\documentclass[]{article}
\usepackage{authblk}
\usepackage{epstopdf}
\usepackage{subfigure}

\usepackage{natbib}
\bibpunct[, ]{(}{)}{;}{a}{}{,}
\renewcommand\bibfont{\fontsize{10}{12}\selectfont}

\usepackage{amssymb}
\usepackage{amsmath}

\usepackage{xcolor}
\usepackage{rotating} 
\usepackage{adjustbox}
\usepackage{hyperref}

\newcommand{\red}{\textcolor{red}}

\newcommand{\green}{\textcolor{teal}}

\newcommand{\bfw}{\boldsymbol{w}}
\newcommand{\bfx}{\boldsymbol{x}}
\newcommand{\bfone}{\boldsymbol{1}}
\newcommand{\bfSigma}{\boldsymbol{\Sigma}}
\newcommand{\A}{\mathcal{A}}

\usepackage{enumitem}

\begin{document}


\title{Market-Implied Sustainability: Insights from Funds' Portfolio Holdings}

\author[a]{R. Giacometti}
\author[a]{G. Torri\thanks{CONTACT G. Torri. Email: gabriele.torri@unibg.it}}
\author[a]{M. Bonomelli}
\author[a]{D. Lauria}
\affil[a]{University of Bergamo, Department of Management, Via dei Caniana, 2, Bergamo, 24127, BG, Italy.}

\date{\today}

\maketitle

\begin{abstract}
This paper proposes a framework to construct Market-Implied Sustainability (MIS) scores for individual firms by exploiting fund-level sustainability classifications and granular portfolio holdings. The central idea is that the relative over/under-representation of a stock in sustainability-oriented funds reveals a market-based assessment of its sustainability profile. We implement the methodology in the European context using the Sustainable Finance Disclosure Regulation (SFDR), considering Article 9 (``dark green'') funds as the sustainability-oriented segment and comparing their portfolio compositions to those of other funds.
	
We compute MIS scores for a large cross-section of European companies over the period 2010--2025. We then examine how MIS relates to traditional firm-level ESG ratings provided by LSEG and analyze the determinants of potential divergences between the two measures. Finally, we assess the economic relevance of MIS through portfolio-tilting strategies, ranging from rule-based reallocations to constrained optimal allocation frameworks.
	
The results show that MIS scores capture dimensions of sustainability that differ systematically from conventional ESG ratings. In portfolio applications, tilting toward firms with high MIS scores improves risk-adjusted performance, whereas strategies based solely on ESG ratings do not deliver comparable gains. Overall, the findings suggest that market-implied sustainability measures provide complementary information to fundamentals-based ESG metrics and have practical relevance for asset allocation and regulatory monitoring.

\textbf{keywords:} Sustainable finance; Financial regulation; Investment strategy; quantile regression; Risk-adjusted performance

\end{abstract}

\section{Introduction}

Sustainable investing has become a central theme in modern finance: by directing capital towards investments labelled as ``sustainable'' or ``green'', investors attempt to support the transition to a more sustainable economy. Research on Socially Responsible Investments (SRI) dates back to the late 1980s \citep[e.g.,][]{bruyn1987field,hylton1992socially} and has since broadened into the large and heterogeneous literature on Environmental, Social and Governance (ESG) investing (see, e.g., \citealp{sandberg2009heterogeneity}). The demand for SRI has grown substantially over the last decade (\citealp{EurosifSRI2018,GSIR2020}), driven both by stakeholders' heightened attention to social and environmental issues and by the view that sustainability-related information can be relevant for risk management and long-term returns \citep[e.g.,][]{amel2018and,becchetti2018fishing}.

A key tension in the literature and in practice concerns the object of measurement. On one side we have firm-level ESG scores, produced by specialised data providers, that aim to evaluate companies' exposures and practices using firm-level fundamentals, disclosures, and third-party information \citep[e.g.,][]{billio2021inside,berg2022aggregate}. On the other side, there are several regulatory and voluntary frameworks assess the sustainability of the fund on the basis of the entire asset and risk management process. For example, the Sustainable Finance Disclosure Regulation (SFDR), Principles for Responsible Investment (PRI), or frameworks for Sustainable Development Goals alignment (SDG) classify and evaluate investment products or fund managers according to product-level objectives, processes, and claimed outcomes. These fund-level labels are not direct measurements of the sustainability of the underlying portfolio holdings, but they signal how asset managers position and present their products with respect to sustainability.

This distinction matters. The inclusion (or exclusion) of a firm in a fund labelled as sustainable captures different informational content compared to ESG scores based on firm fundamentals (``fundamentals-based ratings''). Indeed, ESG scores reflect an analyst's or data provider's assessment of a firm's environmental, social and governance quality, whereas the inclusion of an asset in a sustainable portfolio may capture investors' revealed preferences, institutional mandates, or the practical constraints that determine portfolio construction. The possibility that these two information sources diverge raises questions about what is really being measured when investors rely on ESG labels or on fund classifications: are funds labelled as sustainable systematically investing in firms that are highly rated by ESG providers, or do fund-level labels reflect other considerations (strategy type, engagement/impact orientation, exclusion rules, marketing, regulatory positioning)? In addition to this, ESG scores provided by different agencies often diverge, making even more difficult to analyse the relation between sustainability scores and market practices.

Motivated by this tension, the paper proposes and implements a market-based measure of firm sustainability that leverages fund-level labels together with ownership data. Concretely, if funds that explicitly target sustainable investments (for example, SFDR Article 9 ``dark green'' funds) are systematically over-invested (or under-invested) a given stock relative to other funds, that pattern reveals a market-implied signal about the perceived sustainability of that stock. Aggregating these relative portfolio tilts across a wide set of funds produces an implied sustainability score for each company. This approach complements fundamentals-based ESG scores by recovering the implicit assessments embedded in fund managers' portfolio choices and product classifications.

Drawing an analogy with credit risk assessment (where agency ratings coexist with market-based signals) our market-implied approach translates fund-level categorization and portfolio holdings into security-level indicators suitable for comparative analysis and portfolio construction. The method is intentionally agnostic about the internal mechanics used by any one ESG data provider; instead it uses the ``wisdom of fund managers'' as expressed through portfolio composition and regulated product labels.

\subsection{Contributions and goals}

The primary goals of this paper are threefold:
\begin{enumerate}
	\item \textbf{Methodology.} Develop a transparent methodology to extract \emph{Market-Implied Sustainability (MIS)} scores for firms from fund-level sustainability classifications and granular portfolio holdings. We provide a general formal definition of MIS scores and discuss identification and robustness considerations, and we implement the methodology empirically using SFDR fund labels as a concrete and policy-relevant case study.
	\item \textbf{Implementation and comparison.} Implement the MIS indicator empirically using SFDR fund labels and comprehensive ownership data for a large cross-section of European companies, and compare the resulting MIS scores with established firm-level ESG ratings, using ESG scores issued by LSEG as a benchmark. We document similarities and systematic differences across sectors, firm sizes, and ESG dimensions.
	\item \textbf{Financial relevance.} Investigate whether MIS scores contain information relevant for investors by testing portfolio-tilting strategies based on MIS, in comparison with strategies constructed using traditional firm-level ESG scores.
\end{enumerate}

To operationalize these objectives, we formulate the following research hypotheses.

\begin{itemize}
	\item \textbf{H1 (Alignment between fund labels and ESG ratings).} Firms that receive higher fundamentals-based ESG ratings are, on average, more heavily represented in the portfolios of funds classified as sustainable under SFDR (Article 9) than in the portfolios of other funds.
	\item \textbf{H2 (Divergence between market-implied and fundamentals-based measures).} Market-Implied Sustainability (MIS) scores differ systematically from traditional firm-level ESG ratings, reflecting the distinct information content embedded in fund-level classifications and portfolio choices.
	\item \textbf{H3 (Financial relevance of MIS).} Portfolio strategies tilted according to MIS scores exhibit risk and return characteristics that differ from both an untitled benchmark portfolio and portfolios tilted using traditional firm-level ESG ratings.
\end{itemize}

Guided by these goals, the empirical analysis examines MIS scores computed at a quarterly frequency using a sample of approximately 500 European equity funds over the period 2010--2025. Funds are classified according to their SFDR status as reported in 2023. The focus on Article 9 (``dark green'') funds reflects the regulatory relevance and market penetration of the SFDR framework in the European asset management industry.  The SFDR framework, introduced in 2019, has undergone interpretative refinements that resulted in reclassifications of certain funds, particularly within Article 9. To preserve consistency and focus on the most stringent sustainability segment, we adopt the classification in a single period, while explicitly discussing the implications of this retrospective assumption.

Section \ref{sec:ESG} reviews ESG scores and their limitations. Section \ref{sec:SFDR} summarises the SFDR framework and the economic interpretation of fund labels. Section \ref{sec:implied_scores} presents the MIS methodology and the precise research questions and hypotheses. Sections \ref{sec:econometric} and \ref{sec:pfolio_tilting} report the empirical analyses and portfolio experiments, while Section \ref{sec:conclusions} concludes.

\section{Measuring sustainability -- ESG scores} \label{sec:ESG}

ESG (Environmental, Social, and Governance) refers to the three pillars commonly used to assess the sustainability profile of firms. The concept gained prominence following the 2004 ``Who Cares Wins'' report by the United Nations Global Compact \citep{globalcompact2004}, which encouraged the integration of non-financial considerations into investment analysis and asset management. Since then, ESG scores have become a widely used tool for comparing firms along sustainability dimensions, supported by improvements in non-financial reporting \citep[see e.g.][]{benameur2024sustainability} and the expansion of specialized data providers.

ESG metrics are often motivated by their potential financial relevance. Investors view sustainability-related information as a source of risk mitigation and long-term value creation \citep{amel2018and}. However, the empirical evidence on the performance implications of ESG investing remains mixed. Meta-analyses suggest that sustainable and responsible investing is neither systematically inferior nor superior to conventional investing \citep{revelli2015financial,hornuf2024performance}. Some studies emphasize lower downside risk for responsible firms \citep{freeman2010strategic,luo2017social}, while others document higher expected returns for ``sin'' stocks \citep{fabozzi2008sin,hong2009price}. \cite{giese2019foundations} suggests that changes in a company’s ESG characteristics may be a useful financial indicator, providing evidence that ESG information affects valuation and performance through lower costs of capital, higher valuations, and lower exposures to tail risk. In addition, performance may reflect demand effects: \cite{van2021flow} show that flows into ESG funds can affect the prices of ESG stocks. Recent evidence also suggests that ESG controversies may be more informative for performance than aggregate ESG scores \citep{dorfleitner2020esg,elamer2024esg}.

Over the last decade, ESG investing has attracted substantial capital inflows and prompted widespread fund rebranding. More recently, however, the pace of inflows has slowed, and since 2023 closures of sustainable funds have increased \citep{Morningstar2024global}, reflecting evolving investor preferences, the growing politicization of the topic (especially in the U.S., where several Republican-led states enacted laws restricting ESG use in public funds), regulatory complexity, and performance concerns (see \citealp{saetra2024esg}). 

Despite their widespread adoption, ESG scores face well-documented limitations. Rating agencies employ heterogeneous methodologies, weighting schemes, and data sources, leading to substantial disagreement across providers \citep{billio2021inside,berg2022aggregate,lauria2026mean}. Such divergence complicates both empirical analysis and practical implementation. Furthermore, \cite{gibson2021esg} show that disagreement itself may carry pricing implications, suggesting that ESG scores embed both informational and methodological noise, and the meta-analysis by \cite{rubino2026understanding} reports a fragmentation of the literature, in the sense that most research focuses on the consequences of ESG disagreement, while determinants are typically attributed to methodological and reporting differences, underscoring persistent comparability challenges in ESG measurement.

A related strand of the literature investigates whether funds labelled as sustainable deliver portfolio characteristics consistent with their stated objectives, also in relation to the ESG ratings of the holding. \cite{meyers2024sustainable} conduct a matched-pair analysis of sustainable and conventional funds across multiple geographical regions and find that self-labelled sustainable funds exhibit, on average, higher ESG performance and lower sustainability risk than their conventional counterparts, suggesting alignment between labeling and portfolio composition. At the same time, they document substantial heterogeneity across regions, with European portfolios displaying stronger ESG profiles than those invested in North America, Emerging Markets, or Asia. Similar evidence for the U.S. market is provided by \cite{guidolin2024us}, who show that ESG funds tend to hold stocks with higher ESG ratings and lower exposure to ``sin'' industries relative to non-ESG funds, although the differences appear to have narrowed over time, leaving room for partial greenwashing concerns. Likewise, \cite{curtis2021esg} document that ESG mutual funds provide investors with greater ESG exposure and differentiated voting behavior without systematically sacrificing performance.

Taken together, this literature suggests that, while ESG-labelled funds generally differ from conventional funds in economically meaningful ways, inconsistencies and cross-sectional variation persist, and the alignment between labels and underlying portfolio characteristics is neither mechanical nor uniform. These findings motivate complementary approaches that assess sustainability through realized investment behavior rather than relying solely on stated objectives or third-party ratings.

\section{The SFDR framework}\label{sec:SFDR}

While ESG ratings provide firm-level assessments produced by data providers, regulatory frameworks often address sustainability from a product-level perspective, defining how investment funds can disclose and classify their sustainability objectives. In the European context, a central regulatory reference is the Sustainable Finance Disclosure Regulation (SFDR), introduced by the European Commission on 27 November 2019 to harmonize sustainability-related disclosures in the financial services sector.
The SFDR aims to enhance transparency by requiring asset managers and financial advisers to clearly communicate how sustainability risks are integrated into investment decisions, how these risks may affect financial performance, and what impact investments may have on environmental and social outcomes (\url{https://eur-lex.europa.eu/legal-content/EN/TXT/?uri=celex%3A32019R2088}). The regulation applies to a broad set of financial products, including portfolios managed on a discretionary basis, Alternative Investment Funds (AIFs), Insurance-Based Investment Products (IBIPs), pension products and schemes, Pan-European Personal Pension Products (PEPPs), and Undertakings for Collective Investment in Transferable Securities (UCITS). 
The SFDR does not mandate the incorporation of specific sustainability criteria into the investment process; rather, it establishes a disclosure framework that governs how sustainability claims must be substantiated and communicated.

Focusing our attention of  UCITS funds,\footnote{UCITS is a regulatory framework that allows for the sale of cross-boundary mutual funds for EU member states, created so that retail investors have transparent, regulated, and cross-border investment opportunities.} asset managers have to classify their funds according to one of these three classes: 
\begin{itemize}
	\item Article 9: Also known as ``dark green'', the category covers products targeting bespoke sustainable investments and applies ``[...] where a financial product has sustainable investment as its objective''.
	\item Article 8: Funds that promote environmental or social characteristics (light green); a financial product promotes, among other goals, environmental or social targets, or a combination of those, provided that the companies in which the investments are made follow good governance practices. They are also known as ``environmental and socially promoting''.
	\item Article 6: Funds without a sustainability scope.	Article 6 covers funds which do not integrate any kind of sustainability into the investment process and could include stocks currently excluded by ESG funds such as tobacco companies or thermal coal producers. While these will be allowed to continue to be sold in the EU, provided they are clearly labelled as non-sustainable, they may face considerable marketing difficulties when matched against more sustainable funds.
\end{itemize}

In the first years after the implementation of the SFDR, the universe of funds classified as ``light green'' (Article 8) or ``dark green'' (Article 9) continued to evolve, with persistent concerns about green washing and regulatory uncertainty.  \cite{Morningstar2024SFDR}  reports  the assets of Article 8 and Article 9 funds exceeded €6 trillion at the end September 2024. The two fund groups represent a consistent share of the European universe at 59.2\%. Still, ``dark green'' funds represent only 3.3\% of the total asset under management. \cite{scheitza2024sfdr} analysed the characteristics of Article 9 funds, finding that in 2023, two years after the beginning of the implementation, 60\% of the funds followed an impact-oriented strategy, while the remaining 40\% of funds follow a general ESG strategy. In contrast, the group of funds that was initially classified as Article 9, and later downgraded to Article 6, was less focused on impact-oriented strategies (pursued by 26\% of the funds), and the remaining 74\% followed a general ESG approach. 

Recent studies have examined the composition of SFDR-classified funds, particularly those categorized as Article 8 and Article 9. \cite{lambillon2023green} show that inclusion in Article 9 funds is associated with certain sustainability commitments and higher ESG ratings, while also documenting heterogeneity across fund types. \cite{martinez2025greening} find mixed and quantile-dependent determinants of firm inclusion, highlighting asymmetric relationships between environmental performance, controversial activities, and portfolio selection. Overall, this literature suggests that the composition of SFDR funds reflects multiple and sometimes heterogeneous criteria, rather than a single, uniform sustainability standard.

The Sustainable Finance Disclosure Regulation is currently under revision with the aim of addressing implementation shortcomings and improving clarity, transparency, and usability for investors and market participants. The European Commission’s proposal published in November 2025 introduces a streamlined categorisation system for sustainability-related financial products, replacing the current Article 8 and Article 9 designations with a set of three voluntary categories, broadly defined as Sustainable, Transition, and ESG Basics, each with clear criteria for the portion of assets aligned with sustainability objectives. The reform also significantly simplifies disclosure requirements by reducing complexity and aligning SFDR with other regulatory frameworks, such as the Corporate Sustainability Reporting Directive, to eliminate duplicative reporting burdens and improve comparability of sustainability information for investors. These changes, still subject to negotiation by co-legislators, are designed to make sustainability disclosures more accessible, reduce compliance costs for providers, and support investor decision-making in the context of diverse sustainability preferences\footnote{\url{https://eur-lex.europa.eu/legal-content/EN/TXT/?uri=CELEX:52025PC0841}}

\section{MIS -- Market-Implied Sustainability scores} \label{sec:implied_scores}

We propose the Market-Implied Sustainability (MIS) score as a novel indicator able to synthesize the sustainability analysis of funds managers. It relies on comprehensive portfolio ownership data to unveil latent signals about the sustainability analysis of fund manager on the individual assets. In the remaining Sections we will focus in particular on the funds identified as Article 9 (``dark green funds'')  by the SFDR framework, compared to the funds classified as article 8 (``light green funds'') and article 6 (no sustainability concerns).

The core of our approach is to identify which assets are over/under-invested in sustainable funds compared to other funds. In particular, for each quarter we identify two groups of funds: sustainable funds on one side, and funds that have limited or no focus on sustainability on the other. Then, for each asset we compute the percentage of funds in each group with a non-zero exposure in it, and we take the difference. Let $i$ be the index of the $i$-th security, its MIS score is:

\begin{equation}
	\textit{MIS}_i = p_{S,i}-p_{\overline{S},i}\label{eq:impl_score_c}
\end{equation}
where:
\begin{itemize}
	\item $p_{S,i}$ is the  \% of sustainable funds with asset $i$-th in portfolio,
	\item $p_{\overline{S},i}$ is \% of other funds with asset $i$-th in portfolios.
\end{itemize}
Assuming long-only positions, the indicator is bounded between $-1$ and $1$, with positive values denoting assets that are over-represented in the funds labelled as sustainable (hence implicitly more sustainable), and negative values denoting assets that are under-represented in such funds (less sustainable).

Furthermore, we point out that the measure that we provide is relative, and not absolute, as it uses the market funds as a baseline: a score of zero denotes a prevalence of a stock in sustainable funds equivalent to the prevalence in the rest of the market, and is not tied to a specific level of sustainability. As a consequence, an increased level of sustainability across companies, or a rising concern for the sustainability among the managers of non sustainable funds may lead to a change in the baseline, making the comparison of the scores between different time periods problematic.

The indicator is informative when the values are different from zero. Indeed, zero values can be caused by either very small percentages of funds invested in a specific stock, or no difference from the rest of the market. It is therefore relevant to assess the statistical significance of the results, testing if the score is different from zero. Since the score can be interpreted as a comparison of proportions, we can use a standard statistical test. Let  $p_{S,i}$ be the percentage of sustainable funds with asset $i$-th in the portfolio, and $p_{\overline{S},i}$ the percentage in other funds, the test statistic for test the null hypothesis $H_{0}:  p_{S,i}- p_{\overline{S},i}=0$ is:
\begin{equation*}
	Z=\displaystyle\frac{ p_{S,i}- p_{\overline{S},i} }{\sqrt{p(1-p)(\frac{1}{n_{1}}+\frac{1}{n_{2}})}},
\end{equation*}
where $p=\frac{S_{1,i}+S_{2,i}}{n_{1}+n_{2}}$, $n_{1}$ and $n_{2}$ are the number of sustainable and other funds, respectively, and $S_{1,i}$, $S_{2,i}$ are the corresponding number of funds with asset $i$ in their portfolios.\\

In Appendix \ref{sec:alt_SFDR} we present an alternative approach to compute  MIS scores (that we define MISw) that compares the average weights of the exposure to a specific asset among sustainable and the other funds (rather than the percentage of funds with such asset in the portfolio). The advantage of this alternative method is to consider not only the presence of the asset in a fund, but also the relative portfolio weight. This comes at the cost of more noisy information, due to the fact that the asset management strategies may be very diversified across funds, and the effect of outliers would be more marked.

A similar approach to MIS has been used also by \cite{lambillon2023green}, that compute a greenness index of stocks based on the presence of such companies in SFDR Article 9 (dark green) funds. Differently from us, they do not account for the presence of the stocks in funds not labelled as Article 9. This results in a bias toward stocks that are more popular in general, regardless of the sustainability orientation of the fund. Indeed, such approach cannot differentiate between assets chosen by SFDR Article 9 funds due to their good sustainability profile, and the ones chosen due to other factors  such as good risk adjusted performance, or high capitalization. With our indicator instead, a company that is highly capitalized or that is perceived positively by the market is expected to be included in a large percentage of sustainable (specifically Article 9) funds, but if it is equally invested in other less sustainable funds, the score will be close to zero. Similarly, a stock with a smaller capitalization or mediocre financial performance, may be included in a limited percentage of Article 9 funds, but if the percentage is even smaller in other funds, the implied score would be positive, signalling of a high level of sustainability.

Our approach is also related to the one proposed by \cite{van2021flow}, that uses an indicator similar to MISw to study the effect of investment flows to ESG funds on stock prices. Differently from us, he creates two aggregate portfolios, one composed of a set of ESG equity funds (weighted by asset under management), the other by all mutual funds, then computes a revealed-preference measure of investors' ESG taste for a stock by taking the difference of the weights in the two portfolios. Compared to our approach, \cite{van2021flow}, by aggregating the weights of different funds, and considering their asset under management, gives more emphasis on the aggregate exposures, overweighting the role of large funds. The aggregation of portfolio weights makes this approach similar to the alternative measures that we compute in Appendix \ref{sec:alt_SFDR}. Concerning the identification of sustainable funds, the analysis in \cite{van2021flow} identifies sustainable funds based on the presence of ESG-related keywords in the name. Instead, in our empirical analysis will focus on SFDR, using Article 9 funds as the set of sustainable funds.

\section{Construction of MIS Scores Using SFDR Fund Classifications}

In this section, we implement the proposed methodology computing MIS scores using SFDR fund labels as a policy-relevant empirical setting. Specifically, we rely on the second revision of the SFDR classifications issued in 2023 and identify Article 9 (``dark green'') funds as sustainable. The analysis is restricted to European equity funds.

After describing the main characteristics of the dataset, we compute MIS scores and compare them with firm-level ESG ratings provided by LSEG plc, a widely used benchmark in the literature. LSEG ESG scores offer a quantitative measure of companies’ relative sustainability performance based on publicly reported information. Standardized and industry-relevant indicators are aggregated into category scores, which are consolidated into three pillar scores (Environmental, Social, and Governance) and combined into an overall ESG score using industry-specific weighting. The dataset provides broad temporal and cross-sectional coverage, and the scores are widely used by practitioners as well as in academic research. Nevertheless, the documented lack of consistency across ESG data providers (see, e.g., \citealp{billio2021inside,berg2022aggregate}) calls for caution in generalizing the results, and the discussion should therefore be understood as limited to this specific implementation of ESG scores.

The comparison is conducted at two levels. First, at the fund level, we examine whether sustainable and non-sustainable funds differ in terms of the average ESG ratings of their portfolio holdings. Second, at the security level, we analyse linear and non-linear relationships between ESG ratings and MIS scores. Overall, this section addresses Hypothesis H1 (Alignment between fund labels and ESG ratings).

\subsection{Dataset description}

The construction of MIS scores relies on a dataset of historical portfolio holdings for a panel of approximately 500 equity mutual funds and ETFs with a geographic focus on Europe. The portfolio holding data are issued by Lipper and provided by LSEG. The sample covers the period 2010--2025 at a quarterly frequency.

The dataset includes, on average, 358 funds per quarter, although coverage tends to grow over time, starting from 202 funds in 2010 and peaking at 499 in 2023. Looking at the portfolio holdings, we focus only European stocks, which account for 89\% of assets under management, discarding the remainder assets (consisting of stocks from other geographical regions and investments in other asset classes). On average, the funds are jointly exposed each quarter to more than 2,400 European stocks, for a total market value of approximately €154 billion. These figures evolve over time, with the number of distinct assets peaking at the end of 2021 (2,778 names), while total market value increases over the sample period, driven by both investment flows and stock market dynamics. Across the entire sample, we observe exposures to 5,566 unique European stocks.

\begin{figure}[!h]
	\centering
	\includegraphics[width=0.95\textwidth]{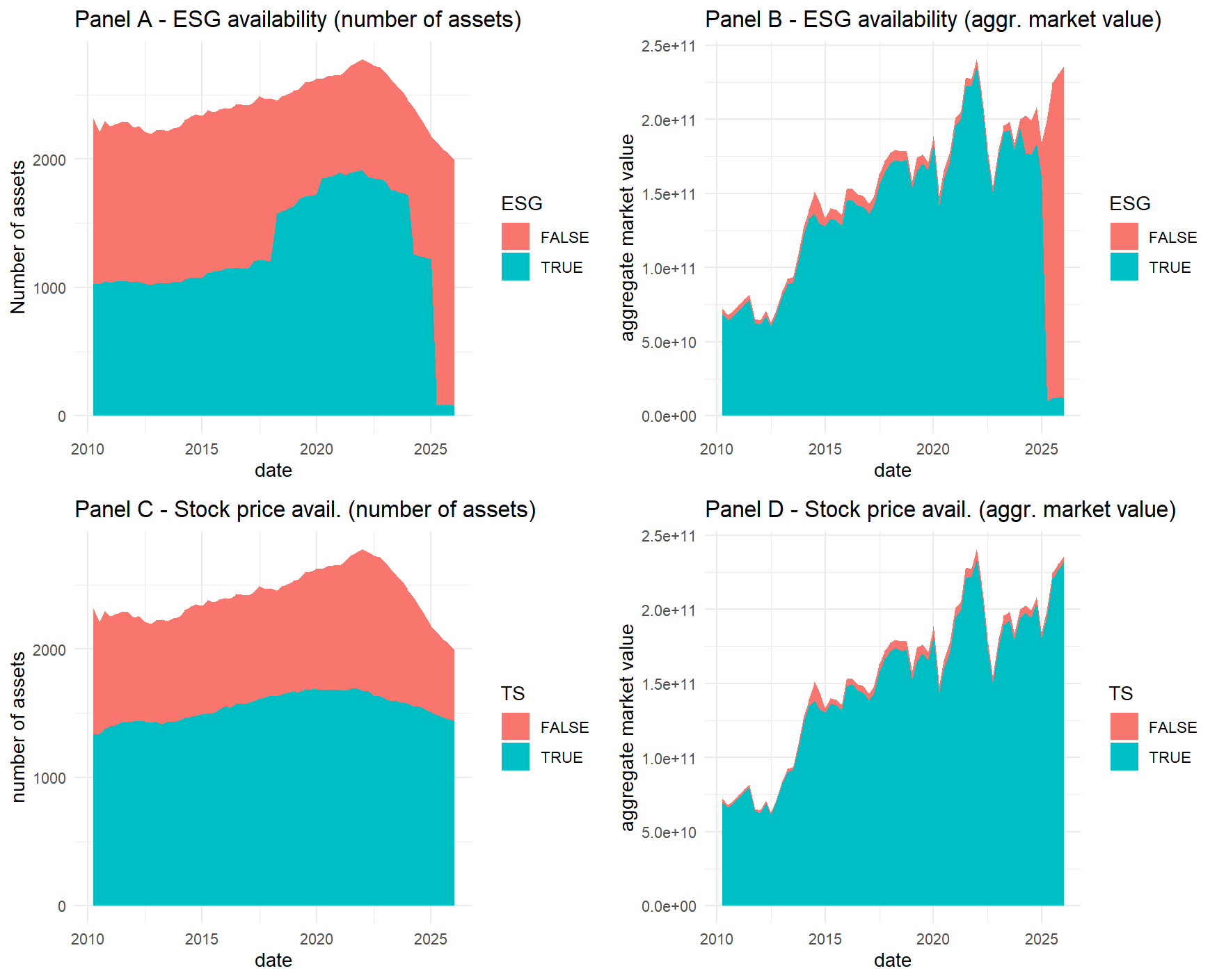}
	\caption{Coverage for LSEG ESG scores (top panels) and daily equity time series (bottom panels) for each quarter. The left panels summarize coverage by number of assets, while the right panels report aggregate market value.}
	\label{fig:descr_coverage}
\end{figure}

\begin{figure}[!h]
	\centering
	\includegraphics[width=0.95\textwidth]{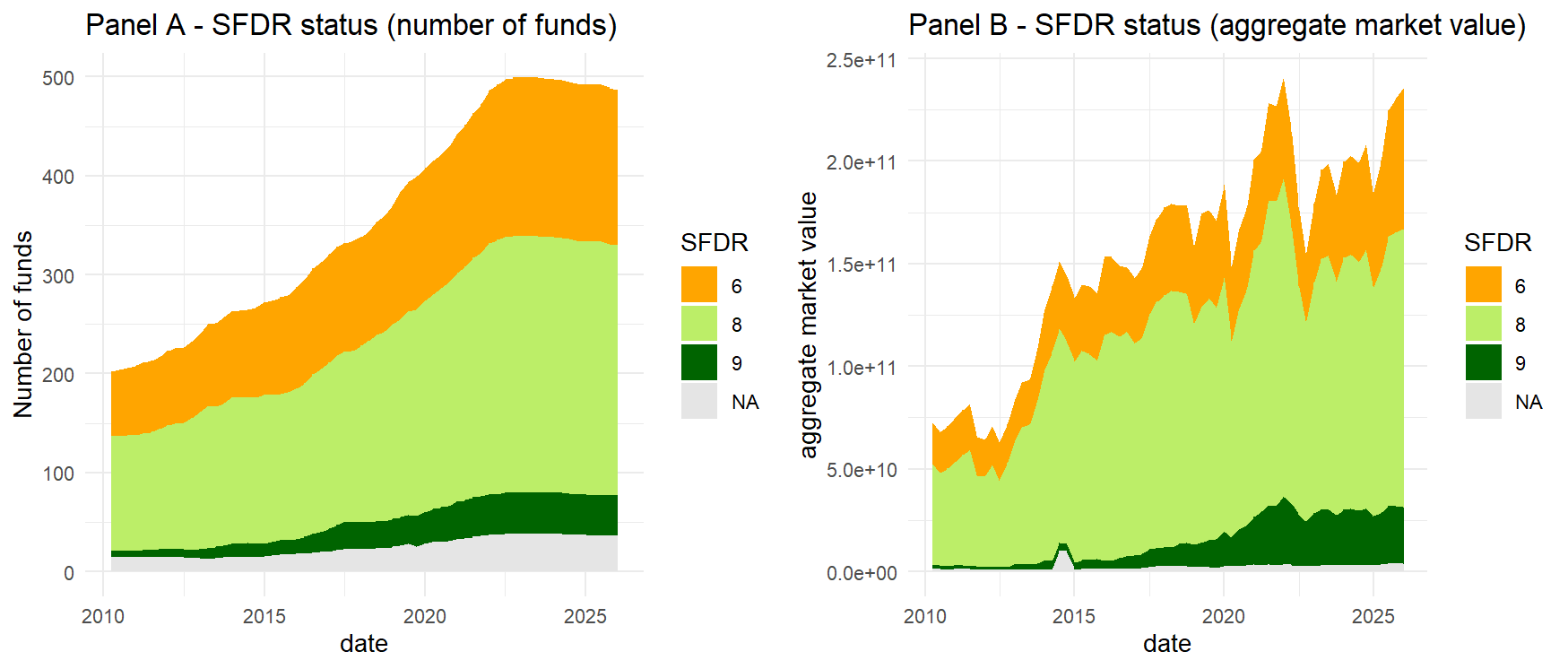}
	\caption{SFDR distribution by quarter. Panel A reports the number of funds by SFDR category (as assigned in 2023) for each quarter. Panel B reports aggregate market value.}
	\label{fig:descr_coverage_SFDR}
\end{figure}

For each stock, we retrieve from LSEG Workspace the daily share price, the annual ESG score, and a set of corporate and sustainability variables at quarterly frequency, which are used in the regression analysis presented in Section \ref{sec:econometric}.

Figure \ref{fig:descr_coverage} illustrates dataset coverage in terms of ESG scores and equity price time series. Coverage is high when measured in aggregate market value (right panels), although ESG scores and price series are missing for a substantial number of smaller stocks (left panels), indicating that the dataset primarily covers firms most commonly included in fund portfolios. We also observe that ESG coverage is limited for 2025, as scores are typically published with a lag in the following year. Therefore, for the econometric analysis in Section \ref{sec:econometric}, we restrict the sample to the period 2010--2024.\footnote{In contrast, for the portfolio analysis in Section \ref{sec:pfolio_tilting}, we use the most recently available information in the market. For example, for portfolios constructed in 2025, we rely on ESG scores for 2024, as they are the latest disclosed figures.}

For each fund in the panel, we assign the corresponding SFDR label (Article 6, Article 8, or Article 9), downloaded from LSEG Workspace. We note that the SFDR framework was introduced in 2019 and that its practical implementation has evolved over time, with subsequent clarifications affecting the classification of certain funds. In particular, part of the initial Article 9 universe was later reclassified following updated supervisory guidance. To ensure consistency in the empirical analysis and to rely on a stable and well-defined set of sustainably oriented funds, we adopt the SFDR labels observed at the end of 2023 and apply this classification throughout the sample period. This choice focuses the analysis on the most stringent segment of the sustainability spectrum. It comes at the cost of a retrospective assumption, which may introduce some bias; however, it avoids embedding transitional reclassifications and evolving regulatory interpretations directly into the time series.

Figure \ref{fig:descr_coverage_SFDR} shows the number of funds in the dataset, together with their distribution across SFDR labels. The majority of funds are classified as Article 8 (``light green''), while only a minority are categorized as Article 9 (``dark green''). The grey segment (labelled ``NA'') represents funds without an available SFDR classification and accounts for a small fraction of the dataset.

\begin{figure}[!h]
	\centering
	\includegraphics[width=\textwidth]{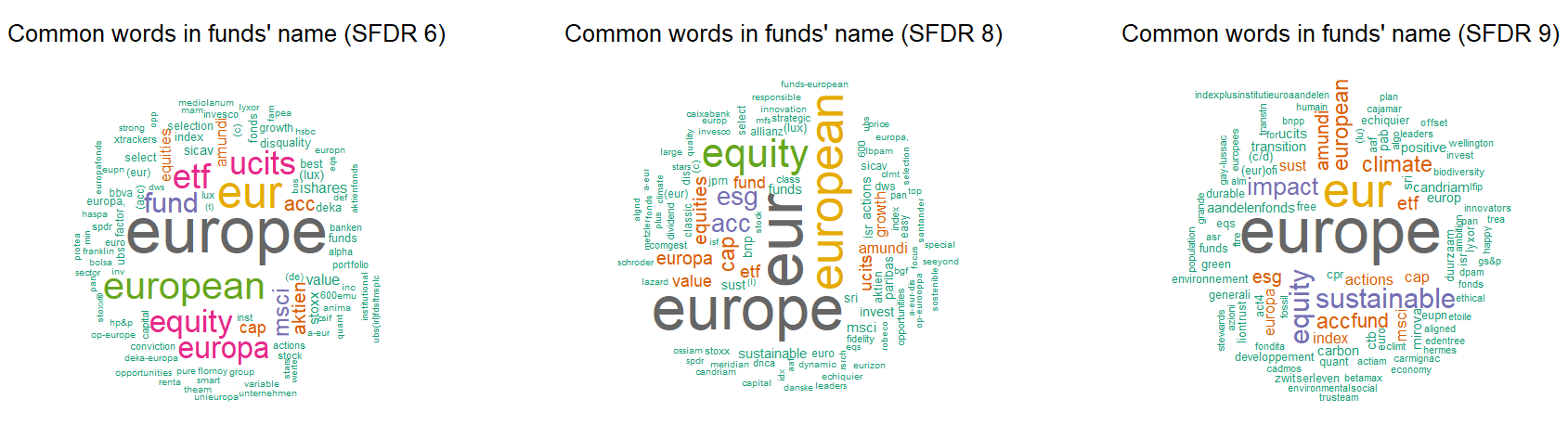}
	\caption{Word cloud of funds' names for SFDR Article 6 (left), 8 (middle), and 9 (right) funds.}
	\label{fig:wordclouds}
\end{figure}

Figure \ref{fig:wordclouds} presents a word cloud summarizing the most frequently used terms in fund names. In addition to generic keywords such as ``Europe'', ``equity'', ``European'', or ``EUR'', Article 9 funds frequently include sustainability-related terms (e.g., ``sustainable'', ``impact'', ``climate'', ``ESG'') in their names. Article 8 funds use such keywords less frequently, with ``ESG'' being the main recurring exception. Article 6 funds, by contrast, rarely include terminology explicitly related to sustainable investing.

\begin{figure}[!h]
	\centering
	\includegraphics[width=0.95\textwidth]{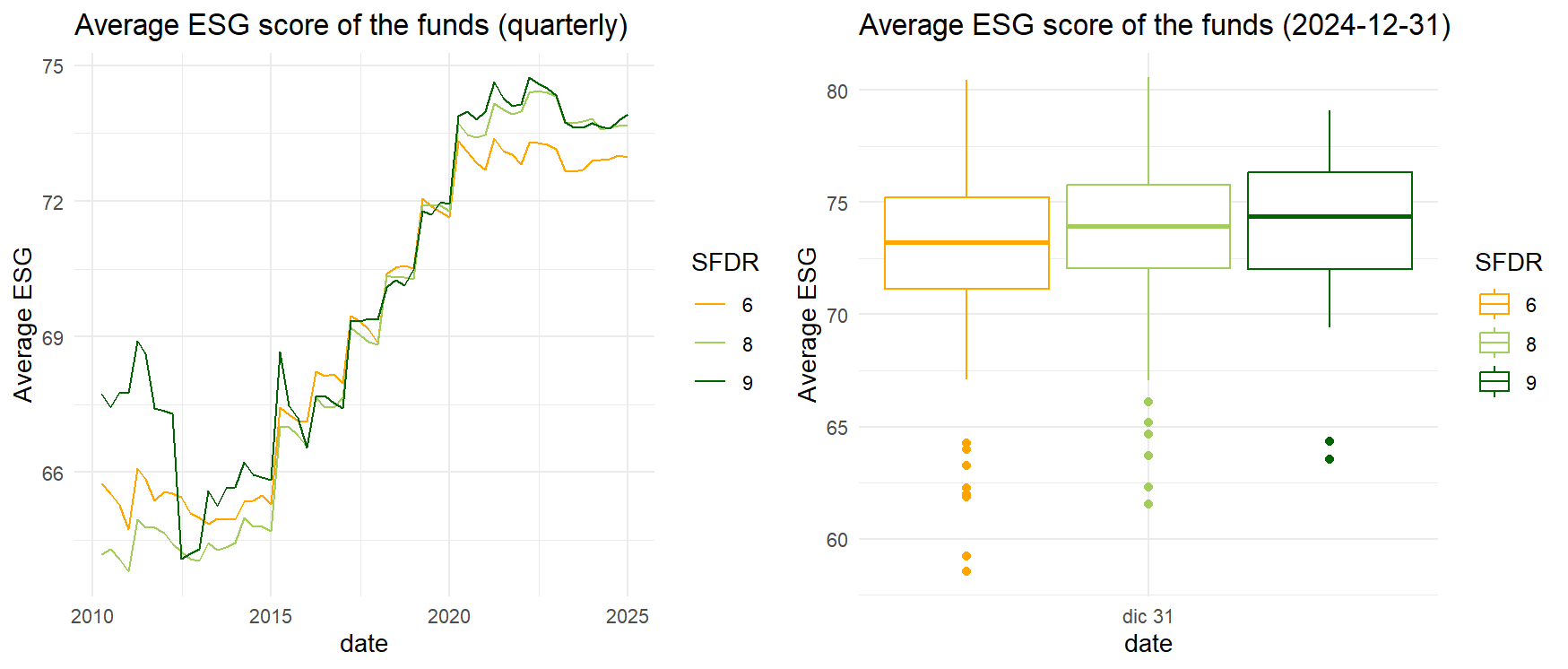}
	\caption{Average fund-level ESG scores by SFDR classification. The left panel shows the time series of average ESG scores for each SFDR class. The right panel displays box plots of the cross-sectional distribution of ESG scores by class as of 31 December 2024.}
	\label{fig:mean_ESG}
\end{figure}

To provide a first assessment of the relationship between SFDR classifications and ESG ratings, we compute the weighted average ESG score of each fund and aggregate the results by SFDR class. This analysis tests Research Hypothesis H1, namely whether Article 9 exhibit systematically higher ESG exposure than other funds.

The left panel of Figure \ref{fig:mean_ESG} reports the time series of average ESG scores by SFDR class. A general upward trend is visible up to 2020, likely reflecting the broader increase in ESG scores provided by LSEG.\footnote{The upward trend in ESG scores issued by LSEG (formerly Refinitiv) has been documented in prior studies and partly attributed to methodological changes in rating practices (see \citealp{benuzzi2023chasing}).} Prior to 2020, differences across SFDR classes are limited and rankings fluctuate over time, with Article 9 funds not consistently displaying higher ESG scores. After 2020, Article 8 and Article 9 funds tend to show moderately higher scores relative to Article 6 funds, although the differences remain contained.

The right panel presents the cross-sectional distribution of fund-level ESG scores in the last quarter of 2024. The distributions largely overlap across SFDR classes, with Article 9 funds exhibiting slightly fewer observations in the lower tail.

Overall, the evidence suggests only partial alignment between SFDR classifications and LSEG ESG scores. This motivates the construction of the MIS score in the next section and the subsequent econometric analysis, which formally investigates the relationship between market-implied and fundamentals-based sustainability measures.

\subsection{MIS and ESG scores} \label{sec:MIS_scores}

Using the methodology outlined in Section \ref{sec:implied_scores}, we compute MIS scores based on SFDR fund classifications. The scores are calculated at a quarterly frequency, consistent with the availability of portfolio holdings data. In line with Research Hypothesis H1, we then compare MIS scores with firm-level ESG ratings provided by LSEG in order to assess the extent of alignment (or divergence) between market-implied and fundamentals-based sustainability measures.

\begin{figure}[!h]
	\centering
	\includegraphics[width=0.95\textwidth]{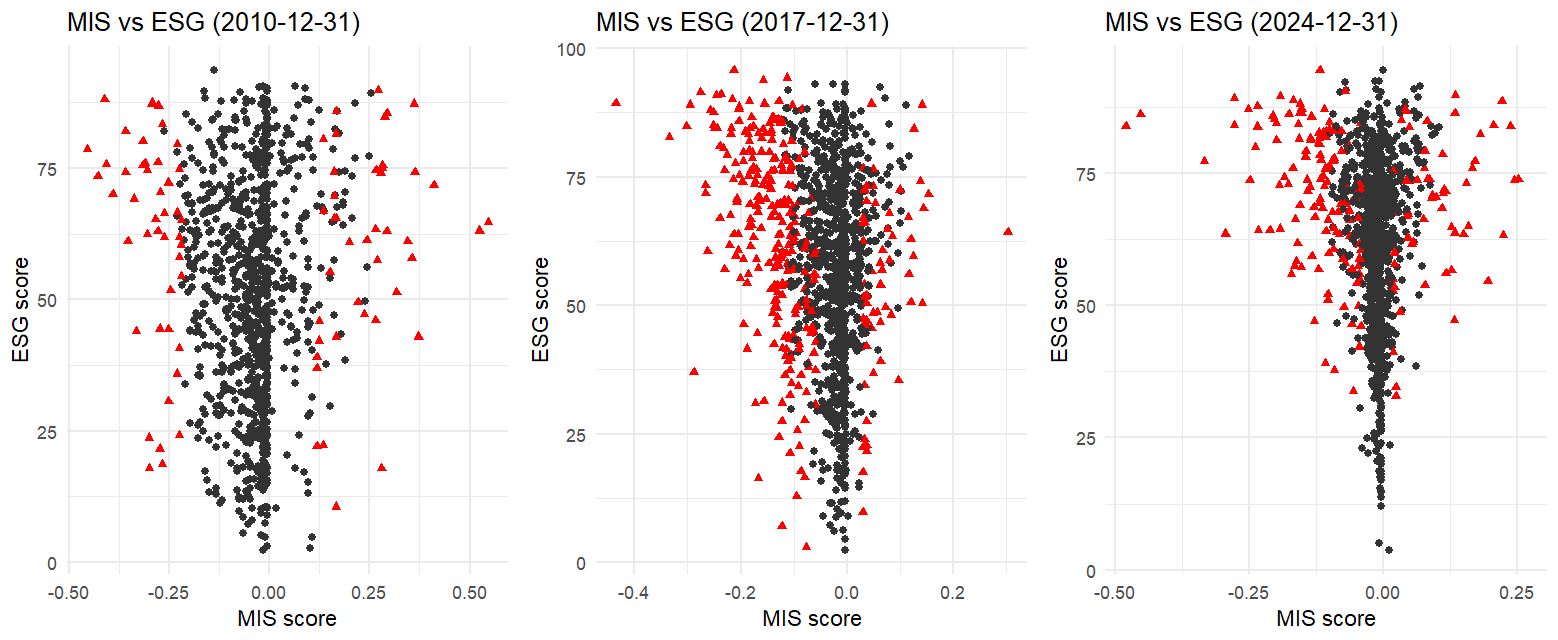}
	\caption{MIS score vs ESG score for three representative time periods. Red dots represent stocks with MIS score statistically significantly different from 0 with 90\% confidence.}
	\label{fig:elephant_ears}
\end{figure}

Figure \ref{fig:elephant_ears} plots MIS scores against ESG ratings for three representative dates. The color scale indicates the statistical significance of MIS scores at the 90\% confidence level. The scatter plots do not display a clear linear association between the two measures. Instead, MIS scores appear increasingly dispersed at higher levels of ESG, particularly in 2024, suggesting a non-linear and asymmetric relationship between the two scoring frameworks. This motivates the use of quantile-based and non-linear regression approaches.

\begin{figure}[!h]
	\centering
	\includegraphics[width=0.95\textwidth]{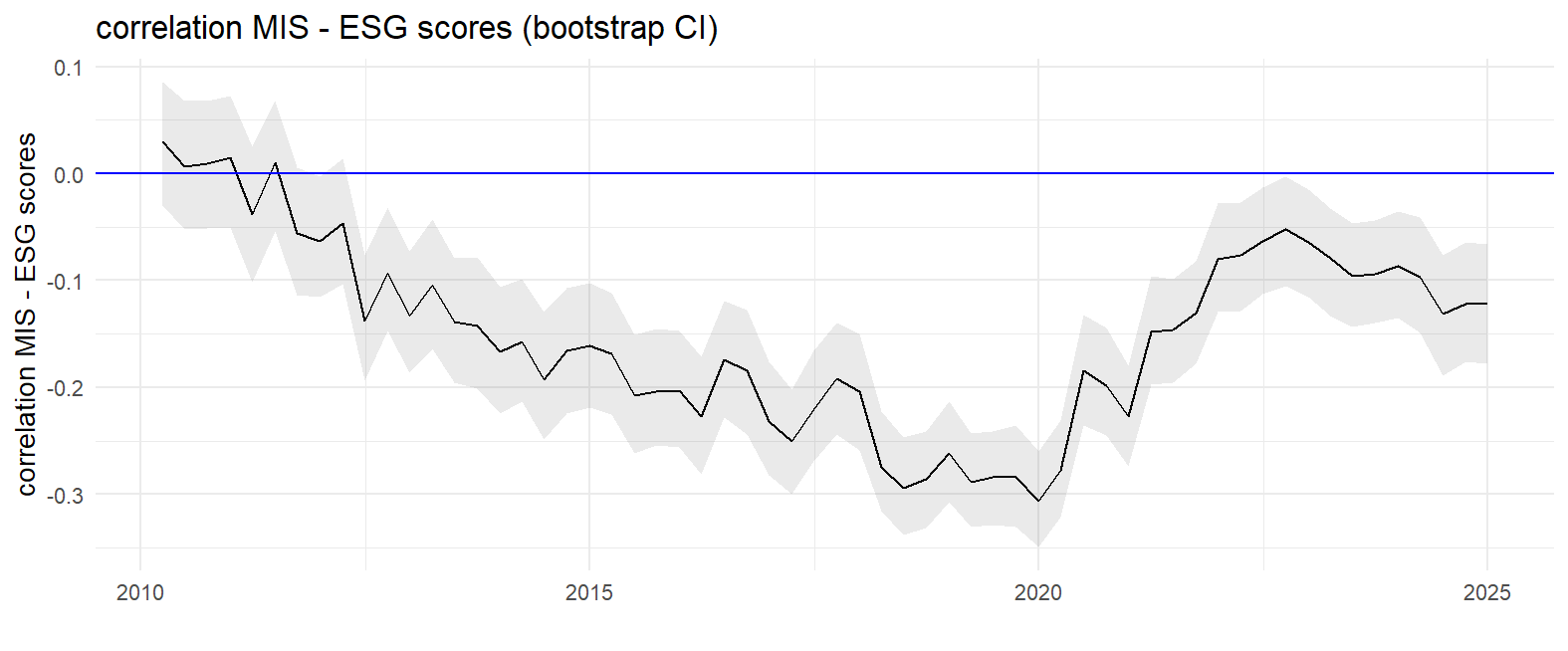}
	\caption{Correlation between MIS scores and ESG scores over time for the period 2010-2024 with bootstrapped 95\% confidence intervals.}
	\label{fig:corr_MIS_ESG}
\end{figure}

Figure \ref{fig:corr_MIS_ESG} reports the Pearson correlation coefficient between MIS and ESG over time with a 95\% bootstrap confidence intervals. The correlation is close to zero at the beginning and end of the sample period and turns negative in the intermediate years. However, the non-linear patterns observed in the bivariate plots suggest caution in interpreting these coefficients, as linear correlation may not adequately capture the dependence structure.

\begin{figure}[!h]
	\centering
	\includegraphics[width=0.89\textwidth]{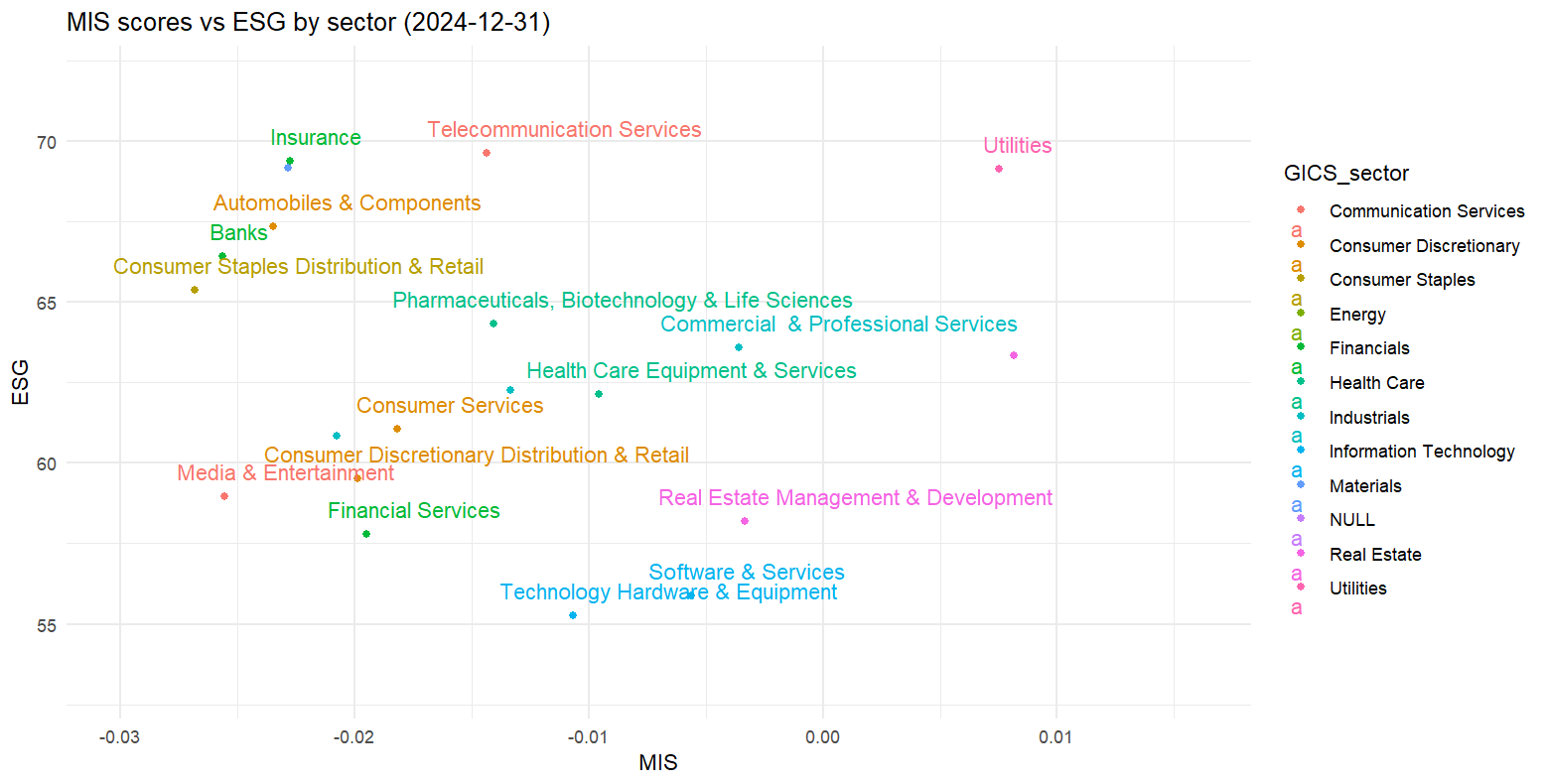}
	\caption{MIS score vs ESG score by sector, computed at 31/12/2024.
	}
	\label{fig:sectors_elephant_ears}
\end{figure}

Considering the sectoral composition of firms, additional patterns emerge. Figure \ref{fig:sectors_elephant_ears} reports the relationship between ESG and MIS scores aggregated at the GICS industry level\footnote{The Global Industry Classification Standard (GICS) is an industry taxonomy developed in 1999 by MSCI and Standard \& Poor's (S\&P) for use by the global financial community}, with colors identifying the broader GICS sector to which each industry belongs. The figure refers to 31 December 2024 as a representative date; results for other quarters (not reported for brevity) display qualitatively similar patterns.

For some industries, the two measures appear broadly aligned. For instance, Utilities exhibit both relatively high average MIS and ESG scores. In contrast, substantial divergences emerge in other segments. Notably, Automobiles \& Components display high ESG ratings but relatively low MIS scores, indicating under-representation in Article 9 funds. A similar pattern is observed for Banks and Insurance companies.

Conversely, industries such as Real Estate Management \& Development, Software \& Services, and Technology Hardware \& Equipment exhibit relatively high MIS scores despite comparatively lower ESG ratings. This suggests that Article 9 funds allocate more heavily to these industries even when their fundamentals-based ESG assessments are less favorable.

\begin{table}[!h]
	\begin{tabular}{l|l|r|r||l|r|r}
		\hline
		\multicolumn{4}{c||}{ranking by MIS score} & 			\multicolumn{3}{c}{ranking by ESG score}\\ \hline
		& name & ESG & implied  & 			name & ESG & implied \\ \hline
 1 &        EDP Renovaveis SA  & 73.9  &  0.253 &           Roche Holding AG  &  94.4 & -0.118 \\
2 &             Prysmian SpA  & 73.6  &  0.245 &            AstraZeneca PLC  &  92.4 & -0.013 \\
3 &    Schneider Electric SE  & 83.8  &  0.238 &        Intesa Sanpaolo SpA  &  92.2 & -0.070 \\
4 &     Dassault Systemes SE  & 63.3  &  0.224 &                  Danone SA  &  91.4 &  0.070 \\
5 &                Alstom SA  & 88.6  &  0.222 &            UPM-Kymmene Oyj  &  90.9 &  0.060 \\
		\multicolumn{4}{c}{...} & 			\multicolumn{3}{c}{...}\\
 n-4 &                 Bayer AG  & 89.2  & -0.278 &      Investment Oresund AB  &  13.7 & -0.002 \\
n-3 &       C. F. Richemont SA  & 63.5  & -0.293 &                Talenom Oyj  &  12.1 & -0.002 \\
n-2 &            Rio Tinto PLC  & 77.2  & -0.333 &               PRS Reit PLC  &  12.0 & -0.002 \\
n-1 &                Shell PLC  & 86.1  & -0.452 &              Kbc Ancora NV  &   5.1 & -0.007 \\
n   &         Total Energies SE  & 83.8  & -0.479 &    Financiere de Tubize SA  &   3.7 &  0.012 \\ \hline
	\end{tabular}
	\caption{Best and worst performers according to MIS score (left) and LSEG ESG score (right). Values computed at 31/12/2024.}\label{tab:portfolio performance}
\end{table}

Table \ref{tab:portfolio performance} reports the names of the companies rated best and worst in terms of MIS score, and  ESG score. We see that the top-performers according to the market based ranking are typically companies involved in the green transition, carrying specific innovations and technologies focused on sustainability. On the contrary, the worst performers are companies involved in controversial sectors such as mining and fossil fuels, or companies that have been hit by controversies in the past. Looking at the ranking by ESG score, we see that the highest ranked companies are large multinational companies belonging to a diversified pool of sectors, while the worst performer are in large part small companies that likely suffer from insufficient sustainability reporting. 

Overall, the results shows how ESG scores and MIS scores carry substantially different information. In particular, it emerges how the market-implied scores allow to identify either companies with a prominent focus on sustainability (positive scores), or companies and sectors that are controversial (negative scores). ESG scores provided by LSEG on the other hand are standardized by sector, and may help investors to pick companies with the best sustainability policies among the peer group of firms. This divergence between the two scoring systems leaves us the question on whether one of the two systems carries relevant information to improve the risk-adjusted performance of a portfolio. In the next section we conduct an analysis to answer such question.

\section{Explaining MIS scores -- a quantile regression analysis}\label{sec:econometric}
In this section we test hypothesis H2, conducting an econometric analysis to identify the determinants of MIS scores computed using SFDR classifications. The selection of explanatory and control variables includes a set of covariates capturing both sustainability-related attributes and firm-level characteristics, together with time and sector fixed effects. To further examine the relationship between MIS and traditional ESG ratings, we incorporate ESG as a key explanatory variable.

Given the non-linear patterns documented in Section \ref{sec:MIS_scores}, we adopt a panel quantile regression framework to allow the effects of covariates to vary across the conditional distribution of MIS scores. Indeed, Figure \ref{fig:elephant_ears} suggests that the relationship between MIS scores and ESG ratings is non-linear and heterogeneous across the distribution, indicating that a standard OLS mean regression may not adequately capture the underlying dependence structure. Specifically, the association appears to differ between low- and high-MIS firms. Formally, we employ the panel quantile regression approach based on the Method of Moments (MMQR) proposed by \cite{machado2019quantiles}. 

The MMQR approach is based on a conditional location--scale model, in the spirit of \cite{he1997quantile} and \cite{zhao2000restricted}. Its main advantage is that it combines the interpretability and tractability of mean-regression panel methods with the flexibility of quantile regression. In particular, it allows the use of methods that are only valid in the estimation of conditional means, such as differencing out individual effects in panel data models, while at the same time providing information on how the regressors affect different parts of the conditional distribution of the dependent variable. This means that instead of estimating only the average effect of a variable (as in OLS), the MMQR approach allows us to understand whether a given factor matters differently for firms with low, medium, or high MIS scores, without sacrificing the ability to control for unobserved firm-specific heterogeneity. MMQR can be adapted to the estimation of cross-sectional models with endogenous variable \citep{machado2019quantiles}. Moreover, it allows individual fixed effects to have heterogeneous effects on the entire conditional distribution of the outcome, rather than constraining their effect to be a location shift only, as in \cite{canay2011simple}, \cite{koenker2004quantile}, and \cite{lamarche2010robust}. Quantile regression is extensively used in the financial literature, including applications that study the composition of sustainable funds, e.g. \cite{martinez2025greening}.

\begin{table}[h!]
{\footnotesize
	\centering
	\caption{Description and source of the dependent variable, main variables, and control variables.}
	\label{tab:variables_description}
	\begin{tabular}{p{3cm}|p{8cm}|p{1.5cm}l}
		\hline
		\textbf{Variables} & \textbf{Description} & \textbf{Source} \\ \hline
		
		\multicolumn{3}{l}{\textbf{Dependent variable}} & \\ \hline
		MIS  score (SFDR-based)& Difference between the percentage of SFDR 9 funds with the $i$-th asset in portfolio and the percentage of other funds (SFDR 6 and SFDR 8) with the $i$-th asset in portfolio. Range: from -1 (worst) to 1 (best). & Computed \\ \hline
		
		\multicolumn{3}{l}{\textbf{Main variable}} & \\ \hline
		ESG score &  ESG score. Range: from 0 (worst) to 100 (best). & LSEG \\ \hline
		
		\multicolumn{3}{l}{\textbf{Sustainability variables}} & \\ \hline
		Green Revenues & Company's green revenue as a proportion of total revenue. & LSEG \\ \hline
		Standardized Total Emission & GHG Emissions,  (Scope 1 + Scope 2) to Revenue USD in million.& LSEG \\ \hline
		Target Reduction & Does the company set specific objectives to be achieved on resource efficiency? (Y/N). & LSEG \\ \hline
		Board Diversity & Proportion of female gender included in the board (\%). & LSEG \\ \hline
		Human Rights Policy  & Does the company have a policy for the exclusion of child, forced or compulsory labour, or to guarantee the freedom of association universally applied independent of local laws? (Y/N). & LSEG \\ \hline
		Armaments & The company produces vehicles, planes, armaments, or any combat materials used by the military (Y/N). & LSEG \\ \hline
		ESG Controversial Score & The company's score in terms of environmental, social, and governance controversies and negative events reflected in global media. Range: from 0 (worst) to 100 (best). & LSEG \\ \hline
		
		\multicolumn{3}{l}{\textbf{Corporate variables}} & \\ \hline
		Size & Total assets (in million €). & LSEG \\ \hline
		P/B Value & Price to book value (\%). & LSEG \\ \hline
		ROE & Return On Equity (\%). & LSEG \\ \hline
		P/E Ratio & Price to Earnings value. & LSEG \\ \hline
		Dividend Yield & Dividend per share as a percentage of the share price (\%). & LSEG \\ \hline
		Sector Dummies & Dummies based on 11 GICS (Global Industry Classification Standard) Sectors. & LSEG \\ \hline 
	\end{tabular}
}
\end{table}

Table \ref{tab:variables_description} provides a description of all variables considered. We sample the variable at quarterly frequency, and we lag the independent variables by one quarter to avoid endogeneity problems and to model delayed effects.

Equation \ref{eq:regression} summarizes the model, in which the conditional quantile of the dependent variable (MIS) is computed as a function of variables that include ESG score of the company, a set of sustainability-related variables, corporate variables, and sector dummies. In order to account for time trends and methodology drift in the ESG scores (see \ref{fig:mean_ESG}), we add time fixed effect to the panel analysis, and we cluster the errors by firm to account for serial correlation.

\begin{equation}
	Q_{MIS_t|X}(\tau) = \beta_{0,	\tau}+\beta_{1,	\tau} ESG_{t-1} + \beta_{2,	\tau} sustainability_{t-1} + \beta_{3,	\tau} corporate_{t-1} + \beta_{4,	\tau} sector. \label{eq:regression}
\end{equation}

For the complete list of regressors we refer to Table \ref{tab:variables_description}. The sustainability-related variables and the control variables have been selected  based on their relevance as key factors influencing the sustainability profile of a company, ensuring a comprehensive and robust analysis. The variables are consistent with the analysis of \cite{lambillon2023green} who examine the characteristics of firms included in SFDR Article 9 funds. In addition, given the strong emphasis posed on the green transition by the European Commission \citep{fetting2020european}, we considered the variable Green Revenues, a metric that reflects companies' exposure to the transition to green economy. We tested for potential collinearity by computing the VIF (Variance Inflation Factor), finding that the values are smaller than 2 for all the regressions performed, suggesting the lack of relevant collinearity in the regressors.

To further enrich the characterization of MIS scores, and to highlight systematic divergence and similarities with traditional ESG scores, we test here the opposite relation compared to Equation \ref{eq:regression}, using the same sustainability and corporate variables, including MIS as a key explanatory variable:

\begin{equation}
	Q_{ESG_t|X}(\tau) = \beta_{0,	\tau}+\beta_{1,	\tau} MIS_{t-1} + \beta_{2,	\tau} sustainability_{t-1} + \beta_{3,	\tau} corporate_{t-1} + \beta_{4,	\tau} sector. \label{eq:regressionESG}
\end{equation}

The analysis is conducted over the period 2010--2024 using quarterly panel data. The year 2025 is excluded, as coverage for ESG scores and other covariates is substantially lower, given that several variables are computed and released only after the end of the reference period.

\subsection{Quantile regression results}

\subsubsection{The determinants of MIS scores} \label{sec:drivers_MIS}

Table \ref{tab:regression_results} reports the main quantile regression results, where the dependent variable is the MIS score. Focusing on the coefficients associated with ESG scores, we observe that the relation is not statistically significant for any of the quantiles, suggesting a lack of alignment between MIS scores and ESG. Still, altough insignificant, we observe that the sign of the coefficient changes from negative to positive at the increase of the quantile, hinting to a positive association between the two frameworks only for firms with relatively high MIS scores, (i.e. securities more strongly represented in SFDR Article 9 funds tend to exhibit higher ESG ratings).

 Turning to the sustainability-related variables, Green Revenues display a positive and statistically significant coefficient for all the quantiles higher or equal than 0.25. This indicates that exposure to green revenues is a robust determinant of MIS and suggests that SFDR Article 9 funds systematically overinvest in firms with higher shares of environmentally aligned activities.The coefficient for ESG controversies exhibit stable signs across quantiles and align with prior expectations, supporting the view that the absence of controversies are important drivers of portfolio selection for Article 9 funds, distinguishing them from the broader fund universe. The coefficients for Standardized Total Emissions and Armaments involvement instead is significant and aligned to expectations only for lowest quantiles, while for higher quantiles the coefficients are not significant. This suggests that these variables are relevant for the exclusion from SFDR Article 9 funds (high emission or involvement with Armaments are associated with low MIS scores), but particularly low emissions or lack of involvement in the production of weapons do not, by themself, guarantee inclusion in dark green funds.
	
In contrast, the coefficients on Target Reduction Policies, Board Diversity, and Human Right Policy shows generally non-significant coefficients and a marked heterogeneity across quantiles, signalling that these specific factors are not among the most relevant determinants of inclusion in Article 9 funds.

Regarding corporate characteristics, firm size (measured by total assets) exhibits a negative and statistically significant coefficient across all quantiles. This indicates that larger firms tend to have lower MIS scores, suggesting that SFDR Article 9 funds are relatively more exposed to smaller firms compared to the broader fund universe.
	
We also find positive and moderately significant coefficients for the price-to-book (PB) and price-to-earnings (PE) ratios. This pattern suggests that firms with higher valuation multiples, typically associated with growth expectations and forward-looking profitability, are more strongly represented in Article 9 funds. In contrast, the coefficients for return on equity (ROE) and dividend yield are negative and, for the second, typically statistically significant. This indicates that firms with stronger current profitability or higher payout ratios are relatively less emphasized, consistent with a tilt toward growth-oriented companies that reinvest earnings rather than distribute them. Overall, these results point to a systematic growth tilt in Article 9 funds, with a preference for smaller, higher-valuation firms over large, mature, income-generating companies.

\begin{table}[h!]
		\centering
		\small
		\caption{Regression baseline results -- determinants of MIS scores.}
		\label{tab:regression_results}
		\begin{adjustbox}{width=\textwidth}
			\begin{tabular}{p{3.5cm}rr r r r r r}
				\hline
				\textbf{Dep. variable: MIS} & \textbf{(1)} & \textbf{(2)} & \textbf{(3)} & \textbf{(4)} & \textbf{(5)} & \textbf{(6)} & \textbf{(7)} \\
VARIABLES & qtile 0.05 & qtile 0.1 & qtile 0.25 & qtile 0.5 & qtile 0.75 & qtile 0.9 & qtile 0.95 \\ \hline
Lag\_ESG\_score & --0.034 & --0.024 & --0.008 & 0.009 & 0.029 & 0.047 & 0.059 \\
& (0.048) & (0.043) & (0.038) & (0.035) & (0.037) & (0.044) & (0.049) \\ \hline
Lag\_green\_revenues & 0.025 & 0.044 & 0.074** & 0.109*** & 0.146*** & 0.182*** & 0.205*** \\
& (0.039) & (0.036) & (0.032) & (0.029) & (0.028) & (0.029) & (0.032) \\
Lag\_std\_total\_emission & --0.002*** & --0.002*** & --0.002*** & --0.002** & --0.001 & --0.001 & --0.001 \\
& (0.001) & (0.001) & (0.001) & (0.001) & (0.001) & (0.001) & (0.001) \\
Lag\_target\_reduction & --0.252 & --0.093 & 0.149 & 0.433 & 0.738 & 1.025 & 1.214 \\
& (0.911) & (0.813) & (0.721) & (0.735) & (0.882) & (1.100) & (1.271) \\
Lag\_board\_diversity & 0.011 & 0.018 & 0.030 & 0.043 & 0.058 & 0.071 & 0.080 \\
& (0.042) & (0.038) & (0.036) & (0.038) & (0.044) & (0.054) & (0.061) \\
Lag\_hum\_rights\_pol & --2.100 & --1.902 & --1.605 & --1.253 & --0.878 & --0.523 & --0.290 \\
& (1.419) & (1.265) & (1.113) & (1.105) & (1.304) & (1.617) & (1.856) \\
Lag\_armaments & --5.487** & --4.667** & --3.430* & --1.971 & --0.410 & 1.066 & 2.031 \\
& (2.472) & (2.240) & (2.066) & (2.170) & (2.606) & (3.199) & (3.613) \\
Lag\_ESG\_controv. & 0.060*** & 0.059*** & 0.057*** & 0.055*** & 0.052*** & 0.050*** & 0.049*** \\
& (0.017) & (0.015) & (0.013) & (0.011) & (0.010) & (0.011) & (0.013) \\ \hline
Lag\_size & --0.003* & --0.003** & --0.003** & --0.003*** & --0.003*** & --0.003** & --0.003** \\
& (0.002) & (0.001) & (0.001) & (0.001) & (0.001) & (0.002) & (0.002) \\
Lag\_PB\_value & 0.338 & 0.358 & 0.387* & 0.421* & 0.458* & 0.493* & 0.516 \\
& (0.248) & (0.235) & (0.225) & (0.230) & (0.254) & (0.291) & (0.320) \\
Lag\_ROE & --0.001 & --0.004 & --0.009 & --0.016 & --0.022 & --0.029 & --0.033 \\
& (0.030) & (0.027) & (0.025) & (0.025) & (0.027) & (0.031) & (0.034) \\
Lag\_PE\_ratio & 0.011* & 0.011* & 0.012** & 0.012** & 0.013** & 0.014** & 0.014* \\
& (0.006) & (0.006) & (0.006) & (0.006) & (0.006) & (0.007) & (0.008) \\
Lag\_div\_yield & --0.491 & --0.499* & --0.512** & --0.527** & --0.543** & --0.558* & --0.567* \\ \hline
& (0.315) & (0.281) & (0.243) & (0.226) & (0.248) & (0.299) & (0.342) \\
Sector Dummies & yes & yes & yes & yes & yes & yes & yes \\ \hline
Time fixed effects & yes & yes & yes & yes & yes & yes & yes \\ \hline
Observations & 11,932 & 11,932 & 11,932 & 11,932 & 11,932 & 11,932 & 11,932 \\
 \hline
				\multicolumn{8}{l}{\scriptsize{Note: Clustered standard errors are reported in parentheses. * $p<0.1$, ** $p<0.05$, *** $p<0.01$.}}
			\end{tabular}
		\end{adjustbox}
\end{table}

\subsubsection{Does MIS explain ESG score?}\label{sec:drivers_ESG}

Table \ref{tab:regression_results_ESG} presents the results of the quantile regressions that explain ESG scores, aimed at further characterize the relation of MIS to ESG scores. The analysis shows that the coefficient for the MIS score is not statistically significant, and it changes signs across quantiles, confirming once again the limited relation between MIS and ESG scores.

Concerning the sustainability variables, we see that Green Revenues have a limited effect on ESG, suggesting that this metric is not tightly related to ESG scores. The same is true for Standardized Total Emissions, that is never significant. This means that, according to the model, lower emissions are not associated to higher ESG scores. This may be explainable by the fact that ESG scores do not use GHG emissions alone, but they elaborate the data, focusing on instance on the comparison with peers, or on the implementation of specific policies to reduce emissions. The coefficients for Target Reduction and Human Rights Policy instead are in line with expectation, showing higher ESG scores across all quantiles for higher values of these variables, suggesting that these measures are integral to the evaluation of a company's ESG score. The coefficients for the Armaments variable and Board Diversity are instead insignificant and have incoherent sign across quantiles, suggesting that these specific sustainability indifcators play a limited role in the determination of ESG scores. Finally, we observe that the coefficient on the Controversies score is negative and statistically significant, consistent with the fact that ESG ratings are constructed separately from controversy indicators and therefore do not directly internalize their effects.

For the corporate variables, the coefficient on Size is positive and statistically significant at the 1\% level for all quantiles with $\tau\geq0.5$. This is consistent with the existing literature \citep{lambillon2023green}, who find that larger firms are more likely to disclose sustainability information and to adopt formal governance and reporting practices, which may mechanically translate into higher ESG ratings. The remaining variables (PB ratio, ROE, PE ratio, and Dividend Yield, are not stable in sign across quantiles and are mostly statistically insignificant. This lack of consistency indicates that ESG ratings are not systematically linked to valuation multiples or short-term profitability measures. Overall, the results suggest that ESG scores are more closely related to firm scale, disclosure capacity, and reporting quality than to market-based growth characteristics or financial performance indicators.

\begin{table}[h!]\centering
		\caption{Quantile Regression Results: Determinants of ESG Scores}
		\label{tab:regression_results_ESG}
		\begin{adjustbox}{width=\textwidth}
			\begin{tabular}{lccccccc}
				\hline
				\textbf{Dep. variable: ESG} & \textbf{(1)} & \textbf{(2)} & \textbf{(3)} & \textbf{(4)} & \textbf{(5)} & \textbf{(6)} & \textbf{(7)} \\
VARIABLES & qtile 0.05 & qtile 0.1 & qtile 0.25 & qtile 0.5 & qtile 0.75 & qtile 0.9 & qtile 0.95 \\ \hline
Lag\_MIS\_SFDR & 0.088 & 0.071 & 0.045 & 0.016 & --0.007 & --0.025 & --0.035 \\
& (0.073) & (0.063) & (0.050) & (0.039) & (0.036) & (0.039) & (0.042) \\ \hline
Lag\_green\_revenues & 0.001 & --0.003 & --0.007 & --0.013 & --0.017 & --0.021 & --0.022 \\
& (0.037) & (0.032) & (0.025) & (0.020) & (0.019) & (0.021) & (0.023) \\
Lag\_std\_total\_emission & 0.000 & 0.000 & 0.000 & 0.000 & --0.000 & --0.000 & --0.000 \\
& (0.001) & (0.001) & (0.001) & (0.001) & (0.001) & (0.001) & (0.001) \\
Lag\_target\_reduction & 7.641*** & 7.315*** & 6.827*** & 6.268*** & 5.825*** & 5.485*** & 5.300*** \\
& (1.471) & (1.283) & (1.048) & (0.890) & (0.889) & (0.965) & (1.029) \\
Lag\_board\_diversity & 0.139 & 0.119 & 0.089 & 0.055 & 0.028 & 0.007 & --0.004 \\
& (0.093) & (0.079) & (0.060) & (0.043) & (0.037) & (0.041) & (0.045) \\
Lag\_hum\_rights\_pol & 13.309*** & 13.176*** & 12.977*** & 12.749*** & 12.568*** & 12.429*** & 12.354*** \\
& (2.072) & (1.764) & (1.369) & (1.106) & (1.130) & (1.288) & (1.411) \\
Lag\_armaments & --0.344 & --0.144 & 0.157 & 0.500 & 0.773 & 0.982 & 1.096 \\
& (5.234) & (4.505) & (3.455) & (2.418) & (1.908) & (1.873) & (2.001) \\
Lag\_ESG\_controv. & --0.116*** & --0.103*** & --0.084*** & --0.061*** & --0.044*** & --0.031*** & --0.023* \\
& (0.026) & (0.022) & (0.017) & (0.012) & (0.011) & (0.011) & (0.012) \\ \hline
Lag\_size & 0.005 & 0.006 & 0.007 & 0.008*** & 0.009*** & 0.009*** & 0.010*** \\
& (0.007) & (0.006) & (0.004) & (0.003) & (0.002) & (0.002) & (0.002) \\
Lag\_PB\_value & --1.566** & --1.272* & --0.830 & --0.324 & 0.077 & 0.384 & 0.551 \\
& (0.743) & (0.678) & (0.590) & (0.502) & (0.450) & (0.419) & (0.414) \\
Lag\_ROE & 0.139** & 0.110* & 0.066 & 0.017 & --0.023 & --0.053 & --0.069* \\
& (0.064) & (0.058) & (0.051) & (0.045) & (0.041) & (0.039) & (0.039) \\
Lag\_PE\_ratio & 0.014 & 0.012 & 0.010 & 0.007 & 0.005 & 0.003 & 0.002 \\
& (0.012) & (0.010) & (0.008) & (0.007) & (0.007) & (0.008) & (0.009) \\
Lag\_div\_yield & 0.347 & 0.380 & 0.429 & 0.486 & 0.531* & 0.565* & 0.584 \\
& (0.624) & (0.536) & (0.419) & (0.324) & (0.304) & (0.330) & (0.356) \\ \hline
Sector Dummies & yes & yes & yes & yes & yes & yes & yes \\ \hline
Time fixed effects & yes & yes & yes & yes & yes & yes & yes \\ \hline
Observations & 11,707 & 11,707 & 11,707 & 11,707 & 11,707 & 11,707 & 11,707 \\
				\hline
				\multicolumn{8}{l}{Note: Clustered standard errors are reported in parentheses. * $p<0.1$, ** $p<0.05$, *** $p<0.01$.}
			\end{tabular}
		\end{adjustbox}
\end{table}

\subsection{Discussion}

The econometric results indicate that MIS scores constructed from SFDR funds labels, and firm-level ESG ratings from LSEG capture related but distinct dimensions of sustainability. The absence of a statistically robust and linear relationship between MIS and ESG, together with the quantile-dependent sign changes, suggests only partial alignment between market-implied and fundamentals-based measures. In particular, ESG ratings do not systematically predict inclusion intensity in Article 9 funds, especially in the lower and middle parts of the MIS distribution.

These findings are related to the analysis by \cite{lambillon2023green}, whose ``greenness'' metric captures the frequency of firm inclusion in Article 9 funds. Our MIS score extends this approach by incorporating relative portfolio weights and benchmarking allocations against non-sustainable funds. In line with their results, we find that sustainability-related commitments and green business exposure (e.g., Green Revenues) play a relevant role in explaining portfolio inclusion. However, our evidence also shows that ESG ratings per se are not a strong or uniform driver of MIS, reinforcing the view that Article 9 composition reflects a broader and more complex decision process than a simple replication of vendor ESG scores. Moreover, unlike their findings, we do not observe significant effects for variables such as Target Reduction or the presence of a Human Rights Policy. A possible explanation is that our methodology explicitly controls for the presence of assets in the portfolios of non--Article 9 funds, thereby capturing the relative allocation decisions of sustainable funds rather than their absolute holdings.

Our results also resonate with \cite{martinez2025greening}, who document mixed and quantile-dependent determinants of firm inclusion across SFDR categories. Their analysis highlights asymmetric relationships between environmental innovation, resource use, and controversial sector exposure, with effects varying across different parts of the distribution. Similarly, we observe that certain variables, such as emissions intensity and armaments exposure, matter primarily in the lower quantiles of MIS, suggesting that exclusionary criteria are more relevant for avoiding low-MIS firms than for guaranteeing high inclusion. Moreover, the increasing importance over time of emissions and ESG controversy is consistent with the growing materiality of climate and reputational risks emphasized in the recent literature.  This is also coherent with the finding of \citep{dorfleitner2020esg}, who reports growing evidence that ESG controversy-related information can be financially relevant and may drive portfolio choices.

The corporate characteristics further suggest that the market-implied signal embedded in Article 9 allocations is not ``sector-agnostic'' or ``style-neutral''. The negative association between firm size and MIS, together with the positive association with valuation multiples and the negative association with dividend yield, points to a systematic tilt toward smaller and more growth-oriented firms. In contrast, performance measures have a limited association with ESG scores, suggesting a more neutral construction, and confirming the results of \cite{garcia2020forecasting}, who reported a limited ability of industry and financial variables in finding small differences in ESG performance among firms. Overall, this pattern is consistent with the idea that sustainability-oriented mandates may interact with investment styles and opportunity sets, and it provides an economic channel through which MIS and ESG can diverge: firm-level ESG ratings tend to correlate with scale and disclosure capacity, while Article 9 allocations appear to reflect a combination of sustainability screening and portfolio tilts.

Finally, the divergence between MIS and ESG is reinforced by the regression results explaining ESG scores. ESG ratings appear more closely associated with firm size, disclosure intensity, and formal policy adoption, whereas MIS is more strongly linked to green revenue exposure, controversy avoidance, and growth-oriented characteristics. Taken together, these findings support the interpretation that fund-level sustainability classifications generate market-implied signals that are complementary, but not reducible, to traditional ESG metrics.

\clearpage

\section{Portfolio tilting} \label{sec:pfolio_tilting}			
In this Section we investigate if we can exploit the  information embedded in our MIS scores to improve the portfolios' financial and sustainability  performance, testing the research hypothesis H3, evaluating the financial relevance of MIS scores. A portfolio “tilt” is an investment strategy that over-weighs a particular investment style. An example would be tilting to small-cap stocks or value stocks that have historically delivered higher returns than the stock market. Over-weighing to a style is always done with the expectation of achieving a higher return or higher risk-adjusted return than the market. In our case we also expect to increase the sustainability of the initial portfolio. In the first part of our analysis,  we tilt a benchmark index by a fixed amount, over-weighting or under-weighting specific  securities according their  sustainability scores. In a second step, we consider more sophisticated allocation strategies in which, instead of fixed weight variation, the optimal portfolio weights are defined as the solutions of constrained optimization problems. These approaches are implemented in a systematic and transparent manner and, although they are more mechanic than real-world asset allocations, taken together they provide a comprehensive assessment of the economic relevance of MIS in applied portfolio settings, spanning simple rule-based reallocations that resemble passive tilting practices to fully optimized strategies that explicitly balance risk and expected return under sustainability constraints.

\begin{figure}
	\centering
	\includegraphics[width=\textwidth]{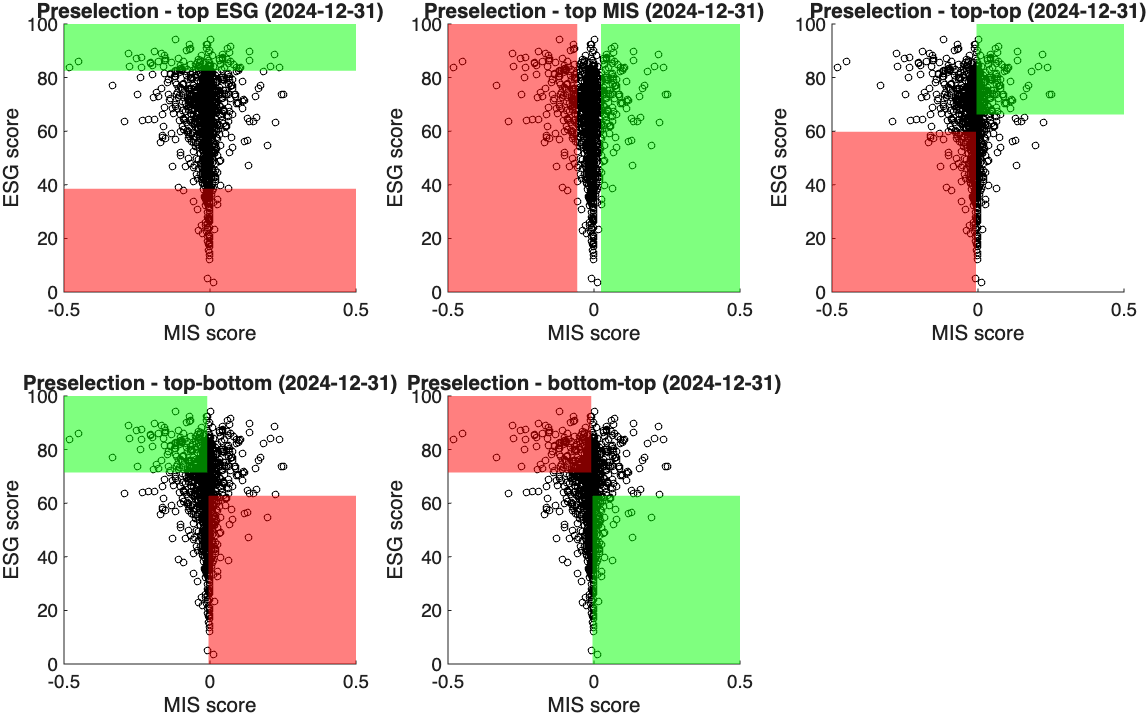}
	\caption{Tilting strategies: the portfolio adds to a baseline index portfolio a long position in the stocks in the green area, and a short position in the red area according to the described procedure. Both the red and the green areas include a number $k=200$ assets.}
	\label{fig:strategies}
\end{figure}

Both the fixed tilting and optimal tilting allocations rely on preselection strategies that identify assets to over/under-weight from an index, on the basis of the ranking of the companies according to the SFDR-based MIS score, and ESG score. The strategies that we consider are  synthetically described below, and illustrated graphically in Figure \ref{fig:strategies}:
\begin{enumerate}[label=$\mathcal{S}$\arabic*]
	\setcounter{enumi}{0}
	\item \textbf{Top ESG} -- over-weight top $k$  companies with best ESG, under-weight $k$ worst ESG.
	\item \textbf{Top MIS} -- over-weight top $k$ companies with best MIS scores, under-weight the $k$ worst  MIS score.
	\item \textbf{Corners TT (Top--Top)} -- over-weight top $k$ highest ESG and MIS scoring companies, under-weight $k$ lowest ESG and MIS.
	\item \textbf{Corners TB (Top ESG--Bottom MIS)} -- over-weight top $k$ highest ESG and lowest MIS, under-weight $k$ lowest ESG and highest MIS.
	\item \textbf{Corners BT (Bottom ESG--Top MIS)} -- over-weight top $k$ highest MIS and lowest ESG, under-weight $k$ lowest MIS and highest ESG.
\end{enumerate}

The interpretation of the first two strategies -- Top ESG and Top MIS -- is straightforward: the assets to over/under weight are selected  based on each of the two scoring systems, respectively. The other strategies aim to assess the interaction effect between ESG and MIS scores. In particular, strategy $\mathcal{S}$3 (``Top--Top'') is an approach that considers and trusts both the market (MIS scores) and the data provider (ESG scores), and over-weights assets rated positively according to both scores, under-weighting the ones with negative scores according to both criteria. Strategies $\mathcal{S}$4 and $\mathcal{S}$5 focus on asset for which ESG and MIS scores diverge, that is, on the groups identified as the ``Top ESG--Bottom MIS'' and the ``Bottom ESG--Top MIS''. In particular strategy $\mathcal{S}$4 over-weights the assets with a high ESG score and low MIS score, and under-weights with a low ESG score and high MIS score.

In practice, to select the assets for strategies $\mathcal{S}$1 and $\mathcal{S}$2 we use the quantiles of the distribution of assets' ESG and MIS scores. For strategies $\mathcal{S}$3, $\mathcal{S}$4, and $\mathcal{S}$5 we use the following approach to identify the groups of assets to under/over-weight using the joint bivariate distribution of the two scores: 
\begin{itemize}
	\item define a grid of probability values $\pi_1, \dots, \pi_m$,
	\item  compute the corresponding  univariate quantiles $Q_{ESG,1}, \dots, Q_{ESG,m}$ and $Q_{MIS,1}, \dots, Q_{MIS,m}$ for both ESG and MIS scores, respectively,
	\item then, to identify the borders of the lower-left quadrant that includes $k$ companies we select the smallest $i$ such that $Pr(ESG<Q_{ESG,i} \land MIS<Q_{MIS,i}) = k/n$, where $n$ is the number of companies. To identify other quadrants we simply invert the sign of the inequalities appropriately.
\end{itemize}

\subsection{Portfolio tilting strategies }

\subsubsection{Portfolio tilting} \label{sec:tilting}
For the tilting strategies, we start from a baseline index portfolio (the STOXX Europe 600), and we add long and short positions based on MIS scores and  ESG scores based on the preselection strategies defined above. We rebalance the portfolio quarterly according to the latest available information.

More in details, each quarter we set up a synthetic index that includes all the assets in the STOXX Europe 600 at the beginning of the quarter, excluding the stocks with incomplete time series. Then on top of it we invest in a long-short portfolio with long positions equal to 10\% of the initial invested amount, and short of the same size. The long and short positions are equally divided between the $k$ best and $k$ worse companies in the preselection strategies, respectively (in this work we consider $k=100$).

\subsubsection{Optimized portfolio tilting} \label{sec:optim}

The analysis of tilting portfolios allows us to assess the aggregate contribution to portfolio performance of ESG and MIS scores. In this section we aim to further analyse the effect on performance, considering active management strategies that incorporate sustainability scores, but where the managers perform stock picking in order to seize opportunities or avoid unnecessary risks.

Funds managers' stock picking strategies are often based on subjective decisions and expert's judgement. To make the analysis reproducible and objective, we thus consider optimal allocation strategies in which the portfolio weights are obtained by solving constrained optimization problems. In particular, we aim to minimize some risk measure or maximise reward-risk ratios while under/over-weighting certain groups of assets compared to an index (e.g. over-weighting the 100 assets with the highest ESG score and under-weighting the ones with the lowest). Formally, we define the set $\A_{\text{under}}$, that includes the best $k$ assets according to a chosen sustainability metric, and $\A_{\text{under}}$ the worst $k$ ones. The under/over-weight is based on the selection strategies discussed previously: ``Top ESG'', ``Top MIS'', ``Top--Top'', ``Top ESG--Bottom MIS'', and ``Bottom ESG--Top MIS''). 

We consider four well know optimal portfolio allocations (mean--CVaR, mean--EVaR, mean--variance, and maximum Sharpe ratio), and we impose constraints on the weights by opportunely add lower and upper bounds to the weights based on their sustainability profile. For the first three strategies (the ones that minimize risk) we constrain the expected return to be greater or equal than the one an index. The four optimal strategies are defined  in Appendix \ref*{sec:model}. We also impose non-negativity constraints.

The models have been implemented in Matlab 2025b using Mosek 10.1 for the linear and quadratic constrained optimizations. The confidence level for the mean--CVaR and mean--EVaR portfolios has been set to $\alpha = 95\%$. 

Since the analysis is focused on the over-weighting and under-weighting of individual positions in a realistic investment setting, we include in the investible universe only the constituents of the index in a given quarter (unlike in \ref{sec:tilting}, where the tilting was based on a broader set of assets). As for the basic tilting, our reference is the STOXX Europe 600, from which we create a synthetic index whose composition is updated quarterly.\footnote{Note that the composition of the synthetic index is marginally different from the one used in \ref{sec:tilting}, as each quarter we include in the index the assets with complete time series and ESG data not only in the quarter, but also in the calibration period required for portfolio optimization.}

\subsection{Empirical set-up}

We run the analysis recalibrating the portfolios every quarter, using the most recent information at the beginning of the investment period to avoid look ahead bias. That is, we use MIS scores computed at the beginning of each period, and the last available published ESG score by LSEG.

For the tilting portfolios in Section \ref{sec:tilting}  we consider $k=100$ (i.e. selecting the best and worst 100 assets for each of the criteria considered), and for each quarter we include in the ranking all the European companies in the sample with complete time series, ESG score, and SFDR-based MIS score.

For the optimized portfolios, we adopt a rolling calibration window of one year (250 daily observations) and a holding period of three months. Portfolios are therefore rebalanced quarterly, with optimal weights estimated using the previous 250 daily returns. We set $k = 100$ (i.e. the 100 best and 100 worse assets are considered for the tilting). The synthetic index is based on composition of the STOXX Europe 600. To ensure implementability and realism, the investable universe is restricted to assets with complete time series and available ESG scores, and membership in the benchmark index at the beginning of each investment period.

Transaction costs are incorporated by applying a one-way cost of 15 basis points to all trades executed at each rebalancing date.

To evaluate whether performance differences are genuinely attributable to sustainability-based selection rather than random allocation effects, we conduct a statistical validation exercise. For each portfolio allocation strategy described in Sections \ref{sec:tilting} and \ref{sec:optim}, we generate 200 alternative portfolios in which assets are randomly selected for over/under-weighting, instead of following the criteria based on MIS or ESG scores. The composition of each random portfolio is determined at the beginning of the sample and kept fixed over time to mimic a stable allocation rule; if an asset exits the sample due to missing data in a subsequent quarter, it is replaced by another asset selected at random.

For each strategy and performance metric, we compute empirical confidence intervals based on the distribution of the 200 randomly generated portfolios. The out-of-sample risk and return measures computed on the portfolios are then compared with the 90\% confidence interval derived from the random allocation benchmark.

\subsection{Analysis of the results}
Tables \ref{tab:portfolio performance_tilting} and  \ref{tab:portfolio performance_opt} report the out-of-sample performance of the tilting portfolios defined in Section \ref{sec:tilting}, and \ref{sec:optim}, respectively. The tables presents annualized return, VaR${95\%}$, CVaR${95\%}$, standard deviation (std), Sharpe ratio, maximum drawdown (maxDD), average turnover, average number of assets (\# assets), maximum individual weight (max $w_i$), and the Herfindahl–Hirschman Index (HHI). Risk and return measures lying outside the $90\%$ confidence interval (CI) obtained from the random allocation benchmark are marked with an asterisk. For ease of interpretation, values outside the confidence interval are highlighted in green when they represent an improvement and in red when they represent a deterioration.

For the simple tilting approach (Table \ref{tab:portfolio performance_tilting}), among the considered strategies, ``Top MIS'' and ``Bottom ESG--Top MIS'' exhibit the strongest out-of-sample performance, characterized by lower risk relative to both the synthetic index and the random portfolios. On the contrary, ``Top ESG'' and ``Top ESG--Bottom MIS'' underperform in terms of both risk and risk-adjusted performance. The differences in risk measures are statistically significant, as they fall outside the confidence intervals derived from the random strategies. In contrast, neither the mean return nor the Sharpe ratio is statistically significant for any strategy. Nevertheless, strategies that overweight firms with high MIS scores (``Top MIS'', ``Top-Top'', and ``Bottom ESG--Top MIS'') compare favorably with the synthetic index in terms of average returns. Overall, the evidence suggests that firms with high MIS scores are associated with superior financial performance, while firms with low MIS scores display weaker risk-return outcomes. In contrast, high-ESG portfolios do not exhibit comparable performance advantages.

The portfolio statistics indicate broadly similar turnover levels across strategies, with ``Top ESG'' displaying the lowest turnover, likely reflecting the greater temporal stability of ESG scores (typically updated annually) compared to MIS scores (computed quarterly). The number of assets, maximum individual exposure, and concentration (HHI) are comparable across strategies.

Table \ref{tab:portfolio performance_opt} reports the performance of the optimized portfolios defined in Section \ref{sec:optim}, under four allocation frameworks (Panel A: mean--CVaR; Panel B: mean--EVaR; Panel C: mean--variance; Panel D: maximum Sharpe ratio) combined with the five preselection strategies.

The results for the three risk-minimization frameworks (Panels A--C) are largely consistent. ``Top MIS'' portfolios combine relatively low risk with returns aligned to the ones of the index, while ``Bottom ESG--Top MIS'' portfolios exhibit low risk but more moderate returns. Strategies overweighting high-ESG firms (``Top ESG'', ``Top--Top'', and ``Top ESG--Bottom MIS'') generally outperform the synthetic index in absolute terms but are often statistically inferior to the corresponding random-allocation portfolios. It is important to note that the confidence intervals are constructed using optimized portfolios with randomly selected over/under-weighted assets, rather than the synthetic index itself, which is typically more volatile than the optimized portfolios.

Under the maximum Sharpe ratio framework (Panel D), ``Top MIS'' and ``Bottom ESG--Top MIS'' again compare favorably, delivering statistically higher Sharpe ratios, higher returns, and lower maximum drawdowns. For ``Bottom ESG--Top MIS'', these gains are achieved at the cost of moderately higher risk.

In terms of portfolio structure, ``Top MIS'' and ``Bottom ESG--Top MIS'' are moderately more concentrated, as reflected in higher HHI values and larger maximum individual weights. They also exhibit higher turnover (approximately 20\%–35\% per quarter). These levels remain economically reasonable given the large number of constituents and the quarterly rebalancing frequency, and all reported performance measures are net of transaction costs.

Taken together, the results suggest that the allocation patterns of SFDR Article 9 (``dark green'') funds embed information that is economically relevant for portfolio performance. This may reflect improved risk management, better identification of transition-related growth opportunities, or lower exposure to controversy-prone firms. By contrast, the ESG scoring framework alone does not appear to be systematically associated with superior performance.

Finally, it is important to emphasize that the strong performance of MIS-based strategies does not imply that all ``dark green'' funds outperform in practice. Fund-level performance depends on additional factors, including sector and geographic allocation, exclusion policies, active stock selection, timing decisions, and management fees. The MIS score aggregates allocation choices across multiple funds and benchmarks them against non-sustainable funds, thereby extracting a market-implied sustainability signal that reflects the collective portfolio decisions of sustainability-oriented investors rather than the performance of any individual fund. The results presented here are therefore complementary to the literature that studies the performance of sustainable funds, such as \cite{becchetti2015socially,hornuf2024performance}.

\begin{sidewaystable}[!h]
{\footnotesize
	\centering
	\begin{tabular}{l|r|r|r|r|r|r|r|r|r|r|r}
		\hline
		\multicolumn{12}{c}{Tilting portfolios}\\ 
		\hline
		tilting portfolios & mean & VaR & EVaR & CVaR & std & Sharpe & maxDD & turnover & \# assets & max $w_i$ & HHI\\ \hline
Synthetic index & {2.72 \%} & {1.64 \%} & {1.27 \%} & {2.6 \%} & {16.36 \%} & {0.17} & {39.23 \%} & {2.53 \%} & {574.17} & {3.03 \%} & {0.01}\\
Top ESG & {2.47 \%} & \red{1.73 \%*} & \red{1.35 \%*} & \red{2.76 \%*} & \red{17.52 \%*} & {0.14} & {41.65 \%} & {9.89 \%} & {670.57} & {3.28 \%} & {0.01}\\
Top MIS & {2.71 \%} & \green{1.59 \%*} & \green{1.23 \%*} & \green{2.49 \%*} & \green{15.83 \%*} & {0.17} & \green{36.21 \%*} & {13.55 \%} & {612.98} & {3.03 \%} & {0.01}\\
Top--Top & {2.98 \%} & {1.64 \%} & {1.26 \%} & {2.57 \%} & {16.28 \%} & {0.18} & {38.63 \%} & {14.69 \%} & {647.11} & {3.1 \%} & {0.01}\\
Top ESG--Bottom MIS & {2.46 \%} & \red{1.7 \%*} & \red{1.33 \%*} & \red{2.73 \%*} & \red{17.25 \%*} & {0.14} & {41.59 \%} & {15.96 \%} & {665.92} & {3.2 \%} & {0.01}\\
Bottom ESG--Top MIS & {2.77 \%} & {1.6 \%} & \green{1.24 \%*} & \green{2.53 \%*} & \green{15.84 \%*} & {0.17} & {37.15 \%} & {15.06 \%} & {664.3} & {2.89 \%} & {0.01}\\

		\hline

	\end{tabular}
	\caption{Out-of-sample portfolio performance of the tilting portfolios. The values with ``*'' denote the measures that are outside the 90\% confidence interval for random selection of assets to under/over-weight. The values outside the confidence interval are green if they are an improvement (e.g. lower risk or higher return), and red if they are a worsening. Synthetic index is the baseline portfolio without tilting.}\label{tab:portfolio performance_tilting}
}
\end{sidewaystable}

\begin{sidewaystable}[!h]
	{\footnotesize
		\centering
		\begin{tabular}{l|r|r|r|r|r|r|r|r|r|r|r|r}
			\hline
			\multicolumn{12}{c}{Panel A -- optimal tilting -- Mean--CVaR portfolios}\\ 
			\hline
			optimal tilting -- CVaR & mean & VaR & EVaR & CVaR & std & Sharpe & maxDD & turnover & \# assets & max $w_i$ & HHI\\
			Synthetic index & {2.68 \%} & \red{1.63 \%*} & \red{1.27 \%*} & \red{2.6 \%*} & \red{16.34 \%*} & {0.16} & {39.51 \%} & {2.39 \%} & {571.68} & {3.04 \%} & {0.01}\\
			Top ESG & {2.65 \%} & \red{1.59 \%*} & \red{1.24 \%*} & \red{2.54 \%*} & \red{15.97 \%*} & {0.17} & \red{39.68 \%*} & {6.04 \%} & {478.44} & {5.02 \%} & {0.01}\\
			Top MIS & {2.72 \%} & \green{1.39 \%*} & \green{1.09 \%*} & \green{2.25 \%*} & \green{14.41 \%*} & {0.19} & {39.39 \%} & {25.17 \%} & {476.62} & {16.99 \%} & {0.05}\\
			Top--Top & {2.57 \%} & \red{1.56 \%*} & \red{1.22 \%*} & \red{2.49 \%*} & \red{15.74 \%*} & {0.16} & {36.4 \%} & {9.88 \%} & {474.95} & {7.09 \%} & {0.01}\\
			Top ESG--Bottom MIS & \red{2.14 \%*} & \red{1.61 \%*} & \red{1.24 \%*} & \red{2.53 \%*} & \red{15.92 \%*} & \red{0.13*} & \red{41.76 \%*} & {7.39 \%} & {476.13} & {4.72 \%} & {0.01}\\
			Bottom ESG--Top MIS & \red{2.09 \%*} & \green{1.43 \%*} & \green{1.11 \%*} & \green{2.28 \%*} & \green{14.29 \%*} & \red{0.15*} & {37.5 \%} & {29.49 \%} & {475.51} & {16.34 \%} & {0.05}\\
			\hline
			
			\multicolumn{12}{c}{Panel B -- optimal tilting -- Mean--EVaR portfolios}\\ 
			\hline
			optimal tilting -- EVaR & mean & VaR & EVaR & CVaR & std & Sharpe & maxDD & turnover & \# assets & max $w_i$ & HHI\\
			Synthetic index & {2.68 \%} & \red{1.63 \%*} & \red{1.27 \%*} & \red{2.6 \%*} & \red{16.34 \%*} & {0.16} & \red{39.51 \%*} & {2.39 \%} & {571.68} & {3.04 \%} & {0.01}\\
			Top ESG & {2.61 \%} & \red{1.61 \%*} & \red{1.25 \%*} & \red{2.56 \%*} & \red{16.08 \%*} & {0.16} & \red{39.7 \%*} & {6.07 \%} & {478.51} & {5.07 \%} & {0.01}\\
			Top MIS & {3.29 \%} & \green{1.36 \%*} & \green{1.09 \%*} & \green{2.24 \%*} & \green{14.33 \%*} & {0.23} & \red{38.8 \%*} & {25 \%} & {477.03} & {16.91 \%} & {0.05}\\
			Top--Top & {2.51 \%} & \red{1.56 \%*} & \red{1.22 \%*} & \red{2.5 \%*} & \red{15.74 \%*} & {0.16} & {38.37 \%} & {9.85 \%} & {475.22} & {6.98 \%} & {0.01}\\
			Top ESG--Bottom MIS & {2.33 \%} & \red{1.6 \%*} & \red{1.25 \%*} & \red{2.55 \%*} & \red{15.99 \%*} & \red{0.15*} & \red{42.23 \%*} & {7.59 \%} & {475.35} & {5.03 \%} & {0.01}\\
			Bottom ESG--Top MIS & {3.03 \%} & \green{1.38 \%*} & \green{1.09 \%*} & \green{2.24 \%*} & \green{13.99 \%*} & {0.22} & {32.82 \%} & {26.88 \%} & {475.25} & {15.32 \%} & {0.04}\\
			\hline
			
			\multicolumn{12}{c}{Panel C -- optimal tilting -- mean--variance portfolios}\\ 
			\hline
			optimal tilting -- MV  & mean & VaR & EVaR & CVaR & std & Sharpe & maxDD & turnover & \# assets & max $w_i$ & HHI\\
			Synthetic index & \red{2.68 \%*} & \red{1.63 \%*} & \red{1.27 \%*} & \red{2.6 \%*} & \red{16.34 \%*} & \red{0.16*} & \red{39.51 \%*} & {2.39 \%} & {571.68} & {3.04 \%} & {0.01}\\
			Top ESG & {2.89 \%} & \red{1.58 \%*} & \red{1.24 \%*} & \red{2.54 \%*} & \red{15.96 \%*} & \red{0.18*} & \red{39.57 \%*} & {5.55 \%} & {480.97} & {4.88 \%} & {0.01}\\
			Top MIS & {2.97 \%} & \green{1.38 \%*} & \green{1.09 \%*} & \green{2.24 \%*} & \green{14.23 \%*} & {0.21} & \red{37.85 \%*} & {22.34 \%} & {479.65} & {16.05 \%} & {0.04}\\
			Top--Top & \red{2.71 \%*} & \red{1.55 \%*} & \red{1.21 \%*} & \red{2.48 \%*} & \red{15.64 \%*} & \red{0.17*} & \red{37.17 \%*} & {9.15 \%} & {477.22} & {6.64 \%} & {0.01}\\
			Top ESG--Bottom MIS & \red{2.78 \%*} & \red{1.58 \%*} & \red{1.23 \%*} & \red{2.52 \%*} & \red{15.87 \%*} & \red{0.18*} & \red{40.84 \%*} & {6.71 \%} & {477.9} & {5.05 \%} & {0.01}\\
			Bottom ESG--Top MIS & \red{2.5 \%*} & \green{1.38 \%*} & \green{1.07 \%*} & \green{2.19 \%*} & \green{13.68 \%*} & \red{0.18*} & \green{32.68 \%*} & {23.45 \%} & {477.44} & {14.88 \%} & {0.04}\\
			\hline

			\multicolumn{12}{c}{Panel D -- optimal tilting -- max--Sharpe portfolios}\\ 
			\hline
			optimal tilting -- max-Sharpe & mean & VaR & EVaR & CVaR & std & Sharpe & maxDD & turnover & \# assets & max $w_i$ & HHI\\
			Synthetic index & {2.68 \%} & \green{1.63 \%*} & \green{1.27 \%*} & {2.6 \%} & \green{16.34 \%*} & {0.16} & {39.51 \%} & {2.39 \%} & {571.68} & {3.04 \%} & {0.01}\\
			Top ESG & \red{2.39 \%*} & \green{1.65 \%*} & {1.28 \%} & {2.61 \%} & \green{16.39 \%*} & \red{0.15*} & {38.93 \%} & {5.99 \%} & {499.33} & {5.08 \%} & {0.01}\\
			Top MIS & \green{5.95 \%*} & {1.69 \%} & {1.29 \%} & {2.64 \%} & \red{17.04 \%*} & \green{0.35*} & \green{32.75 \%*} & {31.08 \%} & {481.94} & {19.33 \%} & {0.06}\\
			Top--Top & {3.83 \%} & \green{1.66 \%*} & {1.29 \%} & {2.64 \%} & {16.68 \%} & {0.23} & {37.51 \%} & {10.27 \%} & {497.43} & {7.44 \%} & {0.01}\\
			Top ESG--Bottom MIS & {3.08 \%} & \green{1.65 \%*} & {1.29 \%} & {2.64 \%} & {16.52 \%} & {0.19} & {40.47 \%} & {7.22 \%} & {497.17} & {5.49 \%} & {0.01}\\
			Bottom ESG--Top MIS & \green{6.11 \%*} & \red{1.85 \%*} & \red{1.38 \%*} & \red{2.79 \%*} & \red{17.88 \%*} & \green{0.34*} & \green{33.09 \%*} & {34.42 \%} & {475.84} & {19.49 \%} & {0.06}\\
			\hline

		\end{tabular}
		\caption{Out-of-sample portfolio performance of the optimal portfolios. The values with ``*'' denote the measures that are outside the 90\% confidence interval for random selection of assets to under/over-weight. The values outside the confidence interval are green if they are an improvement (e.g. lower risk or higher return), and red if they are a worsening.}\label{tab:portfolio performance_opt}
	}
\end{sidewaystable}

\clearpage

\section{Conclusions}\label{sec:conclusions}
	
This paper proposes a replicable framework to construct Market-Implied Sustainability (MIS) scores from fund-level sustainability classifications and granular portfolio holdings. By translating regulated fund labels and observed portfolio constituents into security-level indicators, the approach captures the revealed sustainability preferences of asset managers regarding individual companies. We implement the methodology in the European context using SFDR classifications and a large panel of equity funds, covering the period 2010--2025.

The empirical evidence shows that MIS scores and traditional firm-level ESG scores provided by LSEG reflect different dimensions of sustainability. Firms with high MIS scores are typically associated with activities central to the green transition and with measurable sustainability commitments, such as green revenue, and low exposures to ESG controversies. In contrast, ESG scores display stronger associations with firm size and good sustainability reporting and accountability. The econometric analysis further documents a heterogeneous and non-linear relationship between ESG and MIS across the score distribution, indicating that market-implied assessments do not simply replicate fundamentals-based ratings.

From an investment perspective, MIS scores contain economically relevant information. Portfolio-tilting strategies that overweight equity positions in firms with high MIS scores and underweight ones with low MIS scores improve risk-adjusted performance relative to benchmark allocations, whereas strategies based solely on traditional ESG ratings do not deliver comparable results over the sample period. These findings are robust across both simple tilting rules and constrained optimal allocation strategies. For practitioners, this suggests that market-implied sustainability signals provide a valuable complement to conventional ESG metrics, particularly in environments characterized by disagreement across data providers. For regulators and policymakers, the results highlight that fund-level sustainability classifications have observable implications for capital allocation, and that market-based measures such as MIS can serve as a monitoring tool to assess how regulatory labels translate into portfolio choices.

Taken together, the findings address the three objectives of the paper. We formalize a methodology to derive market-implied sustainability measures; we show empirically that MIS scores differ systematically from established ESG ratings; and we demonstrate that these differences are financially meaningful in portfolio applications. 

The proposed framework admits several extensions. The MIS construction can be applied to alternative sustainability regimes or voluntary standards, enabling cross-jurisdictional comparisons and robustness analyses. It may also be used to detect inconsistencies between fund labels and portfolio composition, contributing to the identification of potential greenwashing practices. More broadly, integrating market-implied signals with fundamentals-based ratings may foster greater convergence and transparency in sustainability measurement, providing both investors and policymakers with a richer understanding of how sustainability considerations are incorporated into financial markets.

\section*{Acknowledgments}
This study was funded by the European Union --  NextGenerationEU, in the framework of the GRINS -- Growing Resilient, INclusive and Sustainable project (GRINS PE00000018 -- CUP F83C22001720001). The views and opinions expressed are solely those of the authors and do not necessarily reflect those of the European Union, nor can the European Union be held responsible for them.

\section*{Disclosure statement}
The authors report there are no competing interests to declare.

\section*{Data availability statement}
The data that support the findings of this study are available from the data provider LSEG. Restrictions apply to the availability of these data, which were used under license for this study.

\clearpage

\appendix

\section{Alternative market-implied sustainability scores}\label{sec:alt_SFDR}

We show here an alternative approach to compute MIS scores. Instead of comparing the percentage of funds with a given asset in the portfolio, we compare the average percentage of portfolio allocated in a given asset:

\begin{equation}
	MISw_i = w_{S,i}-w_{\overline{S},i}\label{eq:impl_score_p}
\end{equation}
where:
\begin{itemize}
	\item $w_{S,i}$ is the  average \% allocation of asset $i$-th  in sustainable funds,
	\item $w_{\overline{S},i}$ is the average \% allocation of asset $i$-th in other funds.
\end{itemize}

We compute the statistical significance of the indicator by considering the standard tests for the comparison of sample means between populations with unequal sample size. The test statistic for test the null hypothesis $H_{0}:  w_{S,i}- w_{\overline{S},i}=0$ is:
\begin{equation*}
	t=\displaystyle\frac{ w_{S,i}- w_{\overline{S},i} }{s_p\cdot \sqrt{1/n_1+1/n_2}},
\end{equation*}
where the pooled standard deviation $s_p$ is
$$s_p=\sqrt{\frac{(n_1-1)s^2_1+(n_2-1)s^2_2}{n_1+n_2-2}},$$

$n_{1}$ and $n_{2}$ are the number of sustainable and other funds, respectively, and the test statistic is distributed as a Student's t distribution with $n_1+n_2-2$ degrees of freedom.

\begin{figure}[!h]
	\centering
	\includegraphics[width=0.95\textwidth]{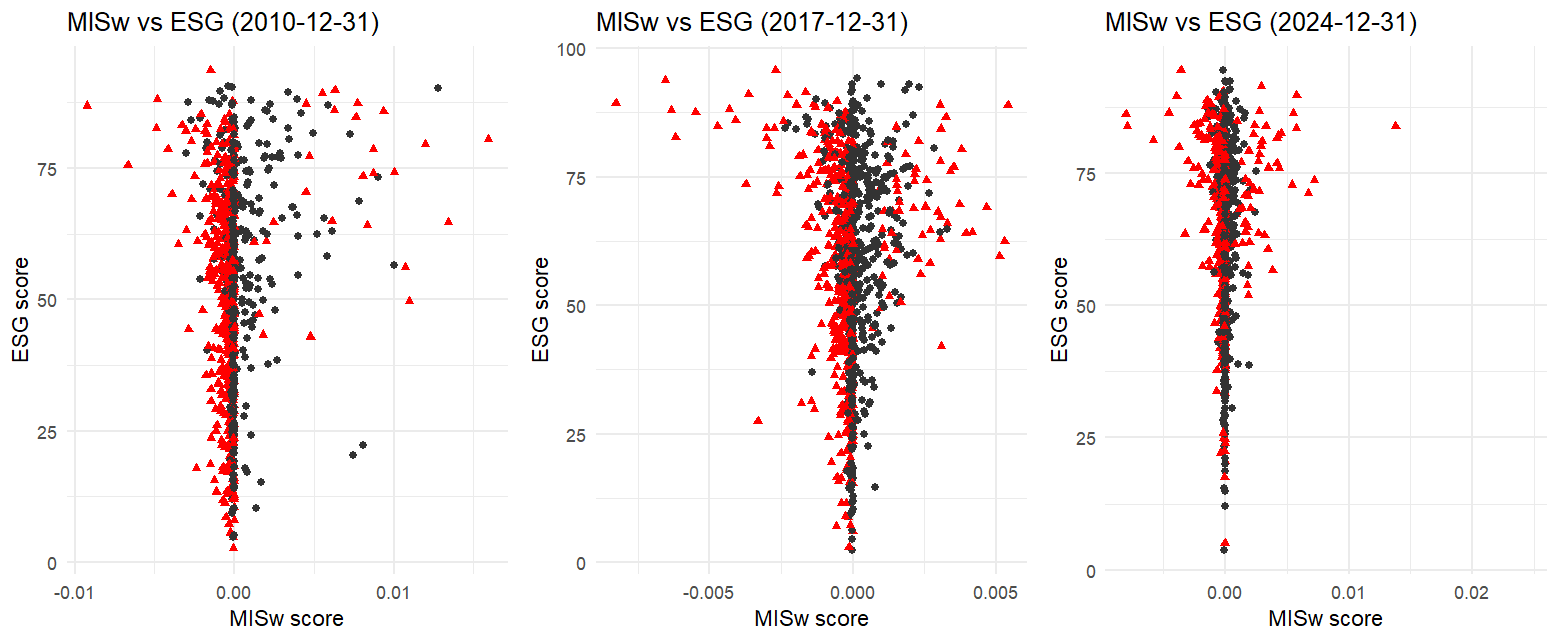}
	\caption{Alternative Market-implied sustainability score (MISw score) vs ESG score for three time periods. Red dots represent stocks with MISw score statistically significantly different from 0 with 90\% confidence.}
	\label{fig:elephant_ears_w}
\end{figure}

\begin{figure}[!h]
	\centering
	\includegraphics[width=0.95\textwidth]{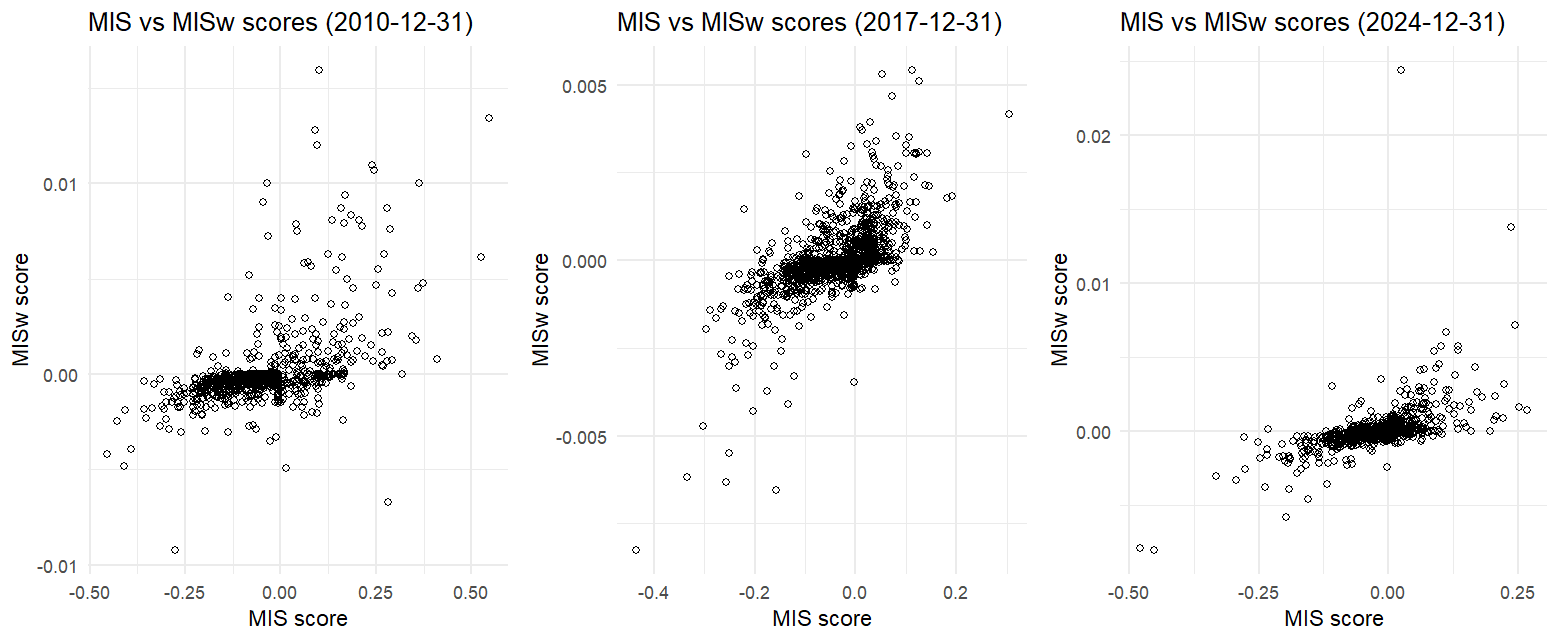}
	\caption{Alternative Market-implied sustainability score (MISw score) vs MIS score for three time periods.}
	\label{fig:SFDR_vs_SFDRw}
\end{figure}

Figure \ref{fig:elephant_ears_w} reports the scatterplot between the alternative MISw score and ESG score. We see that the distributions and the historical evolution are qualitatively similar to the ones in Figure \ref{fig:elephant_ears}. Figure \ref{fig:SFDR_vs_SFDRw} shows the relation between MIS and MISw scores, highlighting a positive relation between the two measures.


\section{Optimisation Models}\label{sec:model}

\begin{itemize}
	\item \textbf{Mean-risk portfolios}. The first three allocation strategies aim to minimize a risk measure $\rho(x)$,\footnote{Denote by $\mathbb{L}^{p}$ the set of all random variables $X$ with $\mathbb{E}[ |X|^{p} ] < +\infty$, defined on a probability space $(\Omega, \mathcal{F}, \mathbb{P})$,
		where $p \in [1, +\infty)$.		 A risk measure is defined as a mapping from a set of random variables to the real numbers  
		$\rho :\mathbb{L}^{p} \to \mathbb {R} \cup \{+\infty \}$.}
	while constraining the expected return to be higher or equal that the one of a reference index portfolio, while respecting the budget constraint and setting upper and lower bounds in order to under/over-weight assets based on the preselection strategies described above:
	
	\begin{align}
		\min_{\bfw\in \mathbb{R}^n} &\rho(R \bfw)\label{eq:opt_man_rho}\\
		s.t. \quad & \bfw'\bfone = 1, \notag\\
		& \mathbb{E}[R]\bfw \geq \mathbb{E}[R_{idx}], \notag\\
		& w_i \leq ub_i,\notag\\
		& w_i \geq lb_i. \notag
	\end{align}
	
	where $\bfw = [w_1, \dots, w_n]'$ is the $[n\times 1]$ vector of portfolio weights, $R$ is the $n-$variate random variable that denotes the returns of the assets, $R_{idx}$ the returns of the index portfolio, and $lb_i$, $ub_i$ the lower and upper bound of asset $i$, computed as follows:
	\begin{equation}
		lb_i = \begin{cases} \label{eq:bounds}
			w_{i,idx} & \text{if } X_i \in \A_{\text{over}},\\
			0 & \text{otherwise}.
		\end{cases}, \quad 
		ub_i = \begin{cases}
			w_{i,idx} & \text{if } X_i \in \A_{\text{under}},\\
			1 & \text{otherwise}.
		\end{cases}
	\end{equation}
	
	The three risk measures considered are variance (mean--variance portfolio),  Conditional Value at Risk (mean--CVaR portfolio), and expectile (mean--EVaR portfolio). The first is the classical Markowitz portfolio \citep{Mark:52} with additional upper and lower bound constraints, and it can be written as a quadratic program with linear constraints. Mean--CVaR portfolio can be formulated as a linear program \citep[see][]{RockUrya00}.	Finally, mean--EVaR is based on the framework in \cite{bellini2021risk}, where the EVaR is the $1-\alpha$ expectile of the portfolio returns' changed in sign \citep{bellini2017risk}. The problem can be expressed as a linear program \citep{bellini2021risk,torri2022penalized}.

	\item \textbf{Maximum Sharpe ratio portfolio}. This strategy maximises the Sharpe ratio of the portfolio, defined as the ratio between excess return and standard deviation. 
	
	\begin{align}
		\max_{\bfw\in \mathbb{R}^n} &\frac{\mathbb{E}[R]\bfw- r_f}{\bfw' \bfSigma \bfw}\label{eq:opt_Sharpe}\\
		s.t. \quad & \bfw'\bfone = 1, \notag\\
		& w_i \leq ub_i,\notag\\
		& w_i \geq lb_i.\notag
	\end{align}

	Assuming that a portfolio with expected return higher that the risk-free rate exists, the problem (\ref{eq:opt_Sharpe}) can be reformulated as a quadratic program:\footnote{With the exception of the upper and lower bound constraints the formulation is analogous to \cite{cornuejols2006optimization}.}
	
	\begin{align}
		\min_{\bfx\in \mathbb{R}^n,y \in \mathbb{R}} &\bfx' \bfSigma \bfx\label{eq:opt_Sharpe_quad}\notag\\
		s.t. \quad &  (\mathbb{E}[R]\bfx- r_f) = 1\notag\\
		& \bfx'\bfone = y\notag\\
		& x_i/y \leq ub_i,\quad i = 1,\dots n\notag\\
		& x_i/y \geq lb_i,\quad i = 1,\dots n.
	\end{align}
	
	where the vector of portfolio weights is $\bfw = \bfx/y$. The problem is quadratic with linear constraints as the last two constraints can be rewritten as $x_i\leq y\cdot ub_i$ and $x_i\geq y\cdot lb_i$, respectively.
	
\end{itemize}
\bibliographystyle{elsarticle-harv} 
\bibliography{BibbaseESGETF}

\end{document}